\documentclass{aa} 

\usepackage{graphicx}
\usepackage{txfonts}
\usepackage{natbib,twoopt}
\usepackage[breaklinks=true]{hyperref} 
\bibpunct{(}{)}{;}{a}{}{,}             

 \usepackage{multirow}

\newcommand{\lya}{Ly$\alpha$}

\def\deg{\hbox{$^\circ$}}

\def\earth{\hbox{$\oplus$}}
\def\lesssim{\mathrel{\hbox{\rlap{\hbox{%
 \lower4pt\hbox{$\sim$}}}\hbox{$<$}}}}
\def\gtrsim{\mathrel{\hbox{\rlap{\hbox{%
 \lower4pt\hbox{$\sim$}}}\hbox{$>$}}}}

\def\arcsec{\hbox{$^{\prime\prime}$}}

\begin{document} 

   \title{High-energy irradiation and mass loss rates\\
          of hot Jupiters in the solar neighborhood}

   \titlerunning{Irradiation and mass loss of hot Jupiters}


   \author{M. Salz\inst{1},
           P. C. Schneider\inst{1},
           S. Czesla\inst{1},
           J. H. M. M. Schmitt\inst{1}
          }

   \institute{Hamburger Sternwarte, Universit\"at Hamburg,
               Gojenbergsweg 112, 21029 Hamburg, Germany\\
              \email{msalz@hs.uni-hamburg.de}
             }

   \date{}

 
   \abstract{ 
     Giant gas planets in close proximity to their host stars experience
     strong irradiation.
     In extreme cases photoevaporation causes a transonic, planetary wind and the 
     persistent mass loss can possibly affect the planetary evolution.
     We have identified nine hot Jupiter systems in the vicinity of the Sun, in which
     expanded planetary atmospheres should be detectable through \lya{} transit spectroscopy
     according to predictions.
     We use X-ray observations with {\it Chandra} and {\it XMM-Newton} of seven of 
     these targets
     to derive the high-energy irradiation
     level of the planetary atmospheres and the resulting mass loss rates.
     We further derive improved \lya{} luminosity estimates for the host stars including interstellar absorption.
     According to our estimates WASP-80\,b, WASP-77\,b, and WASP-43\,b experience the strongest mass loss rates,
     exceeding the mass loss rate of HD\,209458\,b, where an expanded atmosphere
     has been confirmed.
     Furthermore, seven out of nine targets might be amenable to \lya{} transit spectroscopy.
     Finally, we check the possibility of angular momentum transfer from the hot
     Jupiters to the host stars in the three binary systems among our sample,
     but find only weak indications for increased stellar rotation periods of
     WASP-77 and HAT-P-20.
   }


   \maketitle
%

\section{Introduction}

\begin{table*}
\small
\caption{System parameters of the complete sample (see Sect.~\ref{Sect:sample})}             
\label{tabSysPara}      
\centering          
\begin{tabular}{l@{\hspace{-1pt}}l c@{\hspace{6pt}}c@{\hspace{6pt}}c@{\hspace{6pt}}c@{\hspace{6pt}}c@{\hspace{4pt}}r@{}l@{\hspace{6pt}}r@{}l@{\hspace{7pt}}c@{\hspace{3pt}}  l c@{\hspace{6pt}}c@{\hspace{4pt}}c@{\hspace{4pt}}c@{\hspace{7pt}}c@{\hspace{7pt}}c@{\hspace{7pt}}c@{\hspace{6pt}}c }
\hline\hline\vspace{-5pt}\\
    &
    &
  \multicolumn{10}{c}{Host star} & 
    &
  \multicolumn{8}{c}{Planet} \\
  \vspace{-7pt}\\ \cline{3-12}\cline{14-21} \vspace{-5pt}\\
  System & 
    &
  Sp. type & 
  $T_{\mathrm{eff}}$ & 
  $V$ & 
  ($B-V$) & 
  ($J-K$) &
  \multicolumn{2}{c}{$d$} & 
  \multicolumn{2}{c}{$P_{\mathrm{rot}}$} & 
  Age  &
    &
  $R_{\mathrm{pl}}$ & 
  $M_{\mathrm{pl}}$ &  
  $\rho_{\mathrm{pl}}$  &  
  $T_{\mathrm{eq}}$  & 
  $P_{\mathrm{orb}}$  & 
  a  & 
  $e$  &  
  TD \\ 
  \vspace{-9pt}\\
    &
    & 
    & 
  (K)  & 
  (mag) & 
  (mag) & 
  (mag) & 
  \multicolumn{2}{c}{(pc)} & 
  \multicolumn{2}{c}{(d)}  & 
  (Ga)  &
    &
  $(R_{\mathrm{jup}})$ & 
  $(M_{\mathrm{jup}})$ &  
   (g/cm$^{-3}$) & 
   (K) &  
   (d) & 
  (AU) &
    & 
  (\%)  \\
  \vspace{-7pt}\\ \hline\vspace{-5pt}\\ 
  HAT-P-2      &&  F8V   & 6300 &  \hphantom{1}8.7 &  0.46 &  0.19 & 114\,&$\pm\, 10$ &  3.7\,&$\pm\, 0.4$  &  0.4  &&  1.1\hphantom{5}  & \hphantom{1}8.9\hphantom{53}  &  \hphantom{1}7.3\hphantom{3}  &            1700 & \hphantom{1}5.6 & 0.068 & 0.5 & 0.5 \\ 
  \vspace{-8pt}\\
  WASP-38      &&  F8V   & 6200 &  \hphantom{1}9.5 &  0.48 &  0.29 & 110\,&$\pm\, 20$ &  7.5\,&$\pm\, 1.0$  &  1.0  &&  1.1\hphantom{5}  & \hphantom{1}2.7\hphantom{53}  &  \hphantom{1}2.1\hphantom{3}  &            1250 & \hphantom{1}6.9 & 0.076 & 0   & 0.7 \\
  \vspace{-8pt}\\
  WASP-77      &&  G8V   & 5500 &             10.3 &  0.75 &  0.37 &  93\,&$\pm\, 5$  & 15.4\,&$\pm\, 0.4$  &  1.7  &&  1.2\hphantom{5}  & \hphantom{1}1.8\hphantom{53}  &  \hphantom{1}1.3\hphantom{3}  &            1650 & \hphantom{1}1.4 & 0.024 & 0   & 1.7 \\
  \vspace{-8pt}\\
  WASP-10      &&  K5V   & 4700 &             12.7 &  1.15 &  0.62 &  90\,&$\pm\, 20$ & 11.9\,&$\pm\, 0.9$  &  0.6  &&  1.1\hphantom{5}  & \hphantom{1}3.2\hphantom{53}  &  \hphantom{1}3.1\hphantom{3}  & \hphantom{1}950 & \hphantom{1}3.1 & 0.038 & 0   & 2.5 \\
  \vspace{-8pt}\\
  HAT-P-20     &&  K3V   & 4600 &             11.3 &  0.99 &  0.67 &  70\,&$\pm\, 3$  & 14.6\,&$\pm\, 0.9$  &  0.8  &&  0.87             & \hphantom{1}7.3\hphantom{53}  &             13.8\hphantom{3}  & \hphantom{1}950 & \hphantom{1}2.9 & 0.036 & 0   & 1.7 \\
  \vspace{-8pt}\\
  WASP-8       &&  G8V   & 5600 &  \hphantom{1}9.8 &  0.82 &  0.41 &  87\,&$\pm\, 7$  & 16.4\,&$\pm\, 1.0$  &  1.6  &&  1.0\hphantom{5}  & \hphantom{1}2.2\hphantom{53}  &  \hphantom{1}2.6\hphantom{3}  & \hphantom{1}950 & \hphantom{1}8.2 & 0.080 & 0.3 & 1.3 \\
  \vspace{-8pt}\\
  WASP-80      && K7-M0V & 4150 &             11.7 &  0.94 &  0.87 &  60\,&$\pm\, 20$ &  8.1\,&$\pm\, 0.8$  &  0.2  &&  0.95             & \hphantom{1}0.55\hphantom{3}  &  \hphantom{1}0.73             & \hphantom{1}800 & \hphantom{1}3.1 & 0.034 & 0   & 2.9 \\
  \vspace{-6pt}\\
  WASP-43      &&  K7V   & 4400 &             12.4 &  1.00 &  0.73 &  80\,&$\pm\, 30$ & 15.6\,&$\pm\, 0.4$  &  0.8  &&  0.93             & \hphantom{1}1.8\hphantom{53}  &  \hphantom{1}2.9\hphantom{3}  &            1350 & \hphantom{1}0.8 & 0.014 & 0   & 2.6 \\
  \vspace{-8pt}\\
  WASP-18      && F6IV-V & 6400 &  \hphantom{1}9.3 &  0.44 &  0.28 &  99\,&$\pm\, 10$ &  5.0\,&$\pm\, 1.0$  &  0.7  &&  1.3\hphantom{5}  &            10.2\hphantom{53} &             10.3\hphantom{3}   &            2400 & \hphantom{1}0.9 & 0.020 & 0   & 0.9 \\
  \vspace{-6pt}\\
  HD\,209458   &&  G0V   & 6065 &  \hphantom{1}7.7 &  0.58 &  0.28 &  50\,&$\pm\, 2$  &  11.4\,&            &  1.5  &&  1.4\hphantom{5}  & \hphantom{1}0.69\hphantom{3}  &  \hphantom{1}0.34             &            1450 & \hphantom{1}3.5 & 0.047 & 0   & 1.5 \\
  \vspace{-8pt}\\
  HD\,189733   && K0-2V  & 5040 &  \hphantom{1}7.6 &  0.93 &  0.53 &  19\,&           &  12.0\,&$\pm\, 0.1$ &  0.7  &&  1.1\hphantom{5}  & \hphantom{1}1.1\hphantom{53}  &  \hphantom{1}0.96             &            1200 & \hphantom{1}2.2 & 0.031 & 0   & 2.4 \\
    \vspace{-8pt}\\
  55\,Cnc (b)  && K0IV-V & 5200 &  \hphantom{1}6.0 &  0.87 &  0.58 &  12\,&           &  42.7\,&$\pm\, 2.5$ &  6.7  &&   ---             & \hphantom{1}0.80\hphantom{3}  &  ---                          & \hphantom{1}700 &            14.6 & 0.113 & 0   & ---  \\
  \vspace{-8pt}\\                                              
  GJ\,436      && M2.5V  & 3350 &             10.6 &  1.47 &  0.83 &  10\,&           &  56.5\,&            &  6.5  &&  0.38             & \hphantom{1}0.073             &  \hphantom{1}1.7\hphantom{3}  & \hphantom{1}650 & \hphantom{1}2.6 & 0.029 & 0.2 &  0.7 \\
  \vspace{-7pt}\\ \hline                  
\end{tabular}
\tablefoot{Columns are:
           name of the system,
           spectral type,
           effective temperature,
           visual magnitude, 
           colors (SIMBAD or according to spectral type), 
           distance, 
           rotation period, 
           and gyrochronological age of the host star (see Sect.~\ref{Sect:results3});
           planetary radius, 
           mass, 
           density, 
           equilibrium temperature (cited or $T_{\mathrm{eq}} = T_{\mathrm{eff}}\sqrt{R_{\mathrm{st}}/2a}$), 
           orbital period, 
           semimajor axis, 
           orbit eccentricity, 
           and transit depth (TD).
           }
  \tablebib{The data were compiled using exoplanets.org \citep{Wright2011}
            and the following publications:
            HAT-P-2: \citet{Bakos2007, Hipparcos2007, Pal2010}, $P_{\mathrm{rot}}$ from v\,$\sin i$, $T_{\mathrm{eq}}$ varies due to eccentricity (1250 to 2150~K);
            WASP-38: \citet{Barros2011, Brown2012}; 
            WASP-77: \citet{Maxted2013}; 
            WASP-10: \citet{Christian2009, Johnson2009, Smith2009}, $P_{\mathrm{rot}}$ from LSP;
            HAT-P-20: \citet{Bakos2011}, $P_{\mathrm{rot}}$ from LSP; 
            WASP-8: \citet{Queloz2010, Cubillos2012}, $P_{\mathrm{rot}}$ from LSP;
            WASP-80: \citet{Triaud2013}, $P_{\mathrm{rot}}$ from v\,$\sin i$;
            WASP-43: \citet{Hellier2011};
            WASP-18: \citet{Hellier2009, Pillitteri2014}, $P_{\mathrm{rot}}$ from v\,$\sin i$;
            HD\,209458:\citet{Charbonneau2000, Henry2000, Torres2008, SilvaValio2008};
            HD\,189733: \citet{Bouchy2005, Henry2008, Southworth2010};
            55\,Cnc: \citet{Butler1997, McArthur2004, Gray2003, Fischer2008};
            GJ\,436: \citet{Butler2004, Knutson2011}.
           }
\end{table*}

The discovery of giant gas planets in close proximity to their host stars 
brought the stability of these planets and their atmospheres into question.
Orbiting as close as two stellar radii above the photosphere of the host star \citep{Hebb2009}, the so-called hot Jupiters and hot Neptunes are exposed to strong irradiation. In particular, the absorption of extreme ultraviolet (EUV) radiation ionizes hydrogen and heats the atmospheric gas. The resulting high temperatures of about 10\,000~K can support a steady expansion of the atmosphere, which manifests itself in a planetary wind. In its formation this wind is not unlike the solar wind \citep{Parker1958, Watson1981}.
Smaller planets with low densities experience the strongest fractional mass loss and could lose their hydrogen and helium envelopes, evolving to hot Super-Earth like planets \citep{Carter2012}.

Indeed, expanded atmospheres have been confirmed around two hot Jupiters.
In a study of the system HD\,209458, \citet{Vidal2003} discovered a 15\% dimming in the line wings of the hydrogen \lya{} line of the host star when transited by the hot Jupiter in the system, whereas the optical transit depth is only 1.5\% \citep{Henry2000,Charbonneau2000}.
To date, the presence of this upper atmosphere has been confirmed by several observations measuring excess absorption in \ion{H}{i}, \ion{O}{i}, \ion{C}{II}, \ion{Si}{iii,} and \ion{Mg}{i} lines \citep{Vidal2004, Ballester2007, Ehrenreich2008, Linsky2010, Jensen2012, Vidal2013}.
In the second system, HD\,189733, an expanded atmosphere was also confirmed by several observations \citep{Lecavelier2010,Lecavelier2012,Jensen2012,Ben2013}, and the system also exhibits an excess transit depth in X-rays \citep{Poppenhaeger2013}.
Further tentative detections of excess absorption in transit observations of the systems WASP-12 \citep{Fossati2010}, 55\,Cancri \citep{Ehrenreich2012}, and GJ\,436 \citep{Kulow2014} hint that expanded atmospheres could be a common feature in tightly bound gas giants.

Four of these five discoveries succeeded  using \lya{} transit spectroscopy. There are two reasons for this: First, the upper atmospheres of hot gas giants should consist mostly of hydrogen and helium, and second, the \lya{} line dominates the UV spectrum of low mass stars despite interstellar absorption \citep{France2013}.
Nevertheless, with today's instrumentation these types of investigations of exoplanetary atmospheres are only possible in close-by systems with strong \lya{} emission.

Although the presence of expanded atmospheres is commonly accepted at least in the two standard cases, the origin of the \lya{} absorption signal and the essential mass loss are not clear.
Both can be affected by at least three processes, i.e., the planetary wind, the stellar wind, and the stellar radiation pressure.
In general, the spectrally resolved absorption signal reveals the fraction of absorbed stellar emission and the absorption width. The transit depth can be explained by an opaque planetary atmosphere covering a certain fraction of the stellar disk. The absorption width of about 1~\AA{}, corresponding to $\pm 100$~km\,s$^{-1}$, can either be produced by a neutral hydrogen cloud with the given bulk velocity or by a static cloud with sufficient optical depth.

\citet{Koskinen2013a,Koskinen2013b} support the idea that the escaping atmosphere produces a sufficient neutral hydrogen column density to explain the \lya{} transit observations, although this wind proceeds with bulk velocities of only 10~km\,s$^{-1}$.
The transit depth of their model atmosphere is consistent with the measured signal of HD\,209458\,b.
However, the larger velocity offset of $-200$~km\,s$^{-1}$ in the absorption signal of HD\,189733\,b \citep[][]{Lecavelier2012} can probably not be explained with this model.
Among others, \citet[][]{Ekenbaeck2010} argue that charge exchange between the fast ionized stellar wind and the neutral planetary wind creates the hydrogen population causing the \lya{} absorption, but their model requires 40\% of Jupiter's magnetic moment to reproduce the redshifted absorption.
In principle radio observations could be used to determine the magnetic field strength of hot Jupiters, but most host stars have not been detected in radio observations so far \citep{Sirothia2014}. In particular, \citet{Lecavelier2011} presented observations with the Giant Metrewave Radio Telescope resulting in upper limits for the meter wavelength radio emission from HD\,209458 and HD\,189733. With certain assumptions, the nondetections indicate upper limits for the planetary magnetic field strength of only few times that of Jupiter. 
Regarding the field strength  theoretical estimates are not particularly helpful either, as they predict values from insignificant to dominating field strength \citep{Trammell2011}.
Furthermore, neutral hydrogen is also exposed to the radiation pressure from stellar \lya{} emission, which can reach several times the gravitational acceleration of the host star \citep{Bourrier2013}. In a collisionless regime, and with sufficient neutral hydrogen supply, radiation pressure can produce a neutral hydrogen population at velocity offsets of $-100$~km\,s$^{-1  }$. Larger velocity shifts or redshifted absorption cannot be explained by radiation pressure.

Under certain circumstances photoevaporation, charge exchange, and radiation pressure can all affect the atmospheres of hot gaseous planets and especially mass loss rates.
Since each mechanism depends on different system parameters, they should be distinguishable by comparing the absorption signals from different exoplanets. A common parameter of all three processes is the size of the expanded upper atmosphere, which is set by the gravitational potential of the planet and by the EUV irradiation.
A planet of fixed size with a smaller mass produces not only a stronger planetary wind, but also has a larger interaction region for charge exchange and radiation pressure.
Besides the atmospheric size, charge exchange is mainly affected by the temperature and velocity of the stellar wind \citep[][]{Holmstroem2008}, whereas the radiation pressure primarily depends on the \lya{} emission line strength of the host star \citep{Bourrier2013}. While the two systems with secure detections of expanded atmospheres already show differences in their \lya{} absorption signal \citep[e.g., compare][]{Vidal2003,Lecavelier2012}, a larger sample of detections,
increasing the phase space of system parameters, will eventually reveal correlations of the absorption signal with system parameters and help to identify the dominating processes in the atmospheres.

The X-ray luminosity of exoplanet hosts is a crucial parameter for both, estimating the mass loss rate of a hot gas giant and assessing the detectability of the expanded atmosphere.
In the energy-limited case, the mass loss rate depends on the radiative energy input due to hydrogen ionizing emission of the host stars, however, EUV emission is mostly extinguished by interstellar absorption. Several methods exist to reconstruct this emission and particularly the X-ray luminosity is a robust and direct proxy \citep[e.g.,][]{Sanz2011}, because both spectral ranges are formed in associated structures in the stellar atmospheres.
Furthermore, X-rays constitute a significant fraction of the high-energy radiative output from active host stars like HD\,189733 and Corot-2 \citep{Pillitteri2010, Schroeter2011, Sanz2011}.
X-ray luminosities are also closely correlated with the \lya{} luminosity of main-sequence stars \citep{Linsky2013}. 
Thus, they can be used to predict the signal in \lya{} transit observations aimed at a detection of expanded atmospheres.

In an effort to increase the number of planets with detectable expanded atmospheres,
we have identified a sample of nine hot gas giants in the vicinity of the Sun, where predictions yield strong \lya{} emission amenable to transit spectroscopy. 
Here we report on our {\it Chandra} and {\it XMM-Newton} observations of seven targets without prior X-ray observations.
The X-ray observations are used to determine the total high-energy irradiation of the planetary atmospheres. For this purpose, we compare three different methods to reconstruct the EUV emission of host stars.
We present the first energy-limited mass loss rates based on observations.
The improved \lya{} luminosities of our targets in combination with estimates of interstellar absorption reveal the best targets for future transit spectroscopy campaigns.
Finally, we assess the possibility of enhanced stellar rotation induced by tidal interactions with the close gas giants in three binary systems among our sample.


\section{Target selection}
\label{Sect:sample}

To identify the most promising targets for \lya{} transit spectroscopy among confirmed extrasolar planets
we first selected hot, gaseous planets. The recent work by \citet{Marcy2014} has shown that the transition from rocky to gaseous planets with large amounts of volatile elements occurs at $\sim 2~R_{\mathrm{\earth{}}}$.
Therefore, we selected all planets with radii greater than $2~R_{\mathrm{\earth{}}}$ and an orbital distance smaller than 0.1~AU to ensure high levels of irradiation.
Planets with an optical transit depth smaller than 0.5\% were also excluded to increase the contrast of the expected absorption signal.

In a second step, we predicted the strength of the \lya{} emission line, using the following relation between effective temperature $T_{\text{eff}}$, stellar rotation period  $P_{\text{rot}}$, and the \lya{} line flux in erg\,cm$^{-2}$\,s$^{-1}$ at 1~AU \citep{Linsky2013}:
\begin{align}\label{eqLyalpha}
&\log{\left(F_{\text{Ly}\alpha}  (1~\text{AU})\right)} = \\
  &\qquad\begin{cases}
   0.37688 + 0.0002061\, T_{\text{eff}} & \text{for } P_{\text{rot}} = 3-10~\text{d} \, , \\
   0.48243 + 0.0001632\, T_{\text{eff}} & \text{for } P_{\text{rot}} = 10-25~\text{d} \, , \\
   -1.5963 + 0.0004732\, T_{\text{eff}} & \text{for } P_{\text{rot}} > 25~\text{d} \, .
  \end{cases}\notag
\end{align}
The stellar rotation periods were derived from the stellar radius and rotational velocity (see Sect.~\ref{SectPeriods}).
Finally, we scaled the line flux with the stellar distance to obtain the flux at Earth (see Table~\ref{tabSim}).
Interstellar absorption has a decisive effect on the observable strength and profile of the \lya{} line. 
 Because of considerable uncertainty in predicting the absorption,
it is neglected in the selection process, but for the final targets we provide estimates in Sect.~\ref{SectInterstABS}. The distance scaling ensures that the most promising systems are included.

Based on the anticipated \lya{} flux, we ranked the preselected systems to find the most suitable targets for transit spectroscopy campaigns. The ranking order is mostly dominated by the distance of the host stars so that the most promising targets have distances less than 120~pc. Thus, they are probably contained in the Local Bubble of low-density, hot interstellar gas \citep{Redfield2008}. This further limits the uncertainty introduced by interstellar absorption.
Among our final sample we accepted only targets with a \lya{} flux exceeding 1/5 of the unabsorbed line flux from HD\,209458 \citep{Wood2005}. Considering the remaining uncertainty in the prediction process, these targets can exhibit sufficiently strong \lya{} lines for the detection of expanded atmospheres.

With this procedure we found 11 hot gas giants among all confirmed planets. As expected, the best-known systems, HD\,189733 and HD\,209458, lead the ranking. 
The three other systems with reported detections of absorption signals are not to be found in this target list. The distance of WASP-12 is too large for \lya{} transit spectroscopy; 55\,Cnc\,b is not transiting the host star, but only the expanded atmosphere has been found to undergo a grazing transit; and GJ\,436\,b is a slowly rotating host star with no certain detection of the equatorial rotational velocity used to derive estimates for the \lya{} luminosity \citep{Lanotte2014}. 55\,Cnc\,b and GJ\,436\,b have proven to be suitable targets for \lya{} transit observations, thus, we included them in our analysis.
The sample was scanned for previous X-ray observations:
WASP-18 was not detected in a dedicated X-ray observation \citep{Pillitteri2014} and WASP-43 has been analyzed by \citet{Czesla2013}. The remaining seven host stars did not have any prior X-ray observation and unrestrictive {\it ROSAT} upper limits. The system parameters of our targets are summarized in Table~\ref{tabSysPara}.


\subsection{Stellar rotation periods}\label{SectPeriods}

\begin{figure}
  \centering
  \includegraphics[width=1.\hsize, trim=0cm 0cm 0cm 3.2cm, clip=true,]{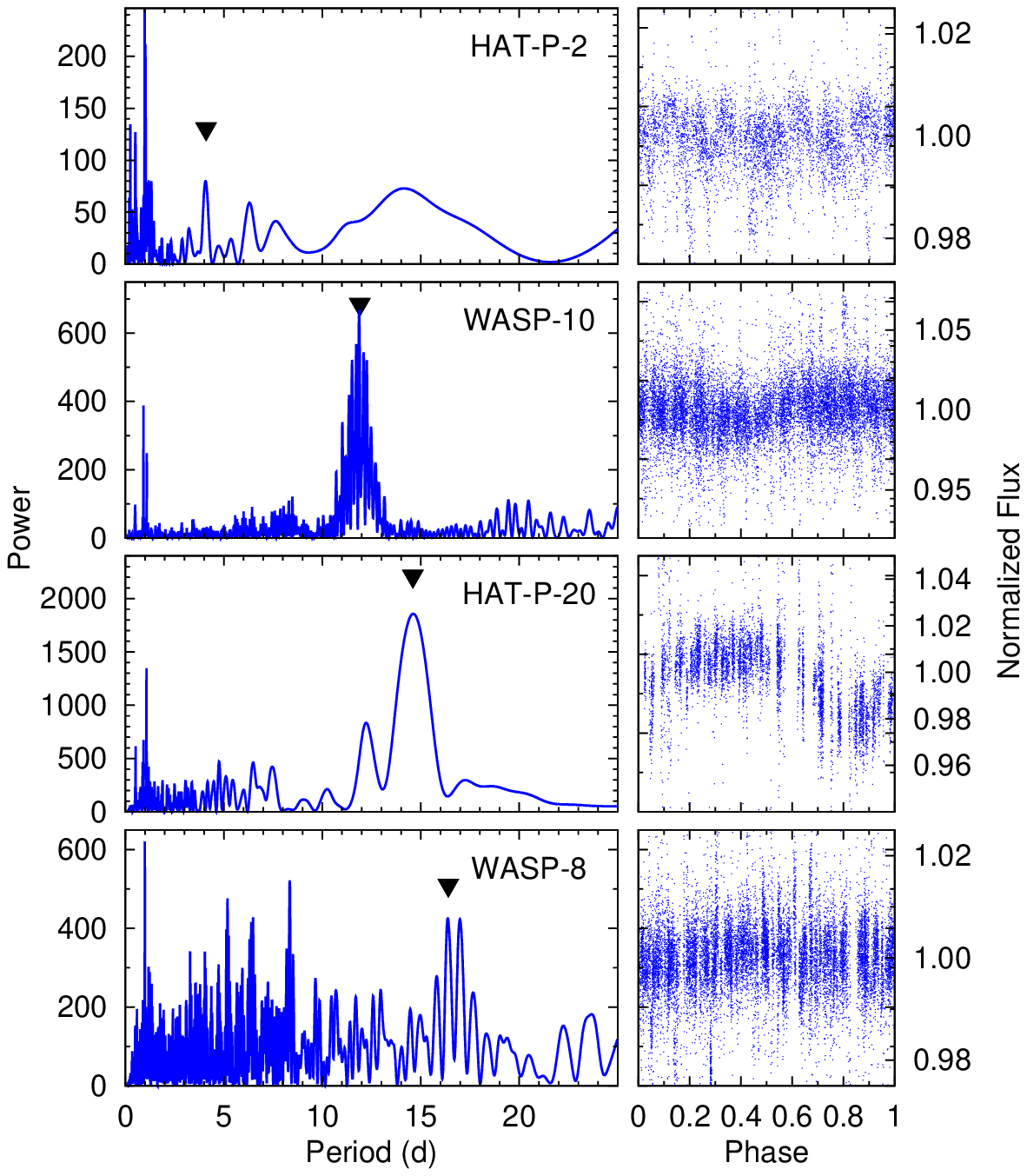}
  \caption{Periodogram of the available stellar photometric data (SuperWASP) and the 
           folded light curves, with the adopted stellar rotation periods
           marked by pointers.
           Periods with a power in excess of 30 occur with a rate of less
           than 1 in 10\,000 by chance. Substructure of major period peaks is
           caused by period or phase shifts between different observing seasons.
           The 1~d period is prominent in all periodograms.
          }
  \label{FigPeriod}
\end{figure}

\begin{table*}
  \caption{Details of the observations}
  \label{tab:obs}
  \centering
  \begin{tabular}{l c c c c c c c}
    \hline\hline
    Target & ObsID & Inst. & Start time & Duration & Start & Stop  & Transit start  \\
           &       &       & (UT)       & (ks)     & phase & phase & phase  \\
    \hline
    HAT-P-2  & 15707      &   C & 2013-11-16 01:15 &  9.9 & 0.971 & 0.996 & 0.984 \\
    Wasp-38  & 15708      &   C & 2014-01-18 07:52 &  9.9 & 0.984 & 1.004 & 0.986 \\
    Wasp-77  & 15709      &   C & 2013-11-09 13:36 &  9.9 & 0.883 & 0.997 & 0.967 \\
    Wasp-10  & 15710      &   C & 2013-11-15 07:31 &  9.9 & 0.786 & 0.830 & 0.985 \\
    Hat-P-20 & 15711      &   C & 2013-11-24 17:33 &  9.9 & 0.699 & 0.749 & 0.987 \\
    Wasp-8   & 15712      &   C & 2013-10-23 20:42 &  9.9 & 0.109 & 0.127 & 0.989 \\
    Wasp-80  & 0744940101 & XMM & 2014-05-15 20:13 & 15.5 & 0.718 & 0.783 & 0.986 \\
    \hline
  \end{tabular}
  \tablefoot{'C' is a {\it Chandra} ACIS-S observation, XMM
             is an {\it XMM-Newton} EPIC PN observation.
             Start and stop phase indicate the exposure duration in reference of the planetary orbit with the transit occurring at phase zero.
            }
\end{table*}

Precise stellar rotation periods of the targets are needed to estimate the \lya{} emission line strength. We compared published values with estimates based on the stellar rotational velocity and our own periodogram analysis of the host star's light curves (SuperWASP \footnote{\url{http://www.superwasp.org/}}).
The adopted values are given in Table~\ref{tabSysPara}.

An initial estimate for the period can be derived from the stellar radius  $R_{\mathrm{s}}$ and the equatorial rotational velocity v$_{\mathrm{eq}}$:
\begin{equation}\label{EqRotation}
  P_\mathrm{rot} = 2\pi R_{\mathrm{s}} \times \mathrm{v}_{\mathrm{eq}}^{-1} \, ,
\end{equation}
assuming that the inclination of the rotation axis is $90^{\circ}$ --- a reasonable assumption for most transiting systems. The rotational velocity and the stellar radius are available for most systems,
and in the cases of WASP-80, HAT-P-2 and WASP-18 more precise rotation periods are not available.

WASP-38, WASP-77, WASP-10, HD\,209458, HD\,189733, 55\,Cnc, and GJ\,436 have published rotation periods based on photometric variability, which correspond well with the values derived from Eq.~\ref{EqRotation}.
The rotation period of WASP-43 was analyzed by \citet{Hellier2011}.

We download the original photometric data of the available targets (HAT-P-2, WASP-10, HAT-P-20, WASP-8) from the SuperWASP archive and analyzed the light curves via a generalized Lomb-Scargle periodogram \citep[LSP,][]{Zechmeister2009} using the PyAstronomy package\footnote{\href{http://www.hs.uni-hamburg.de/DE/Ins/Per/Czesla/PyA/PyA/index.html}{http://www.hs.uni-hamburg.de/DE/Ins/Per/Czesla/PyA/PyA/}}
to derive stellar rotation periods (see Fig.\ref{FigPeriod}).
WASP-10 and HAT-P-20 show isolated peaks with high powers in the periodogram that agree with the v\,$\sin i$ based rotation periods within the errors.
The analysis of HAT-P-2 remained inconclusive, so  we reverted to the rotation period from Eq.~\ref{EqRotation}.

The data of WASP-8 consist of two seasons with different major periods, which have false alarm probabilities of less than $10^{-4}$. Only the period around 16.4~d is present in both seasons and shows a clear photometric variation in the phase folded light curve. The estimate from the stellar rotational velocity and the radius is about a factor of two larger,
but the planet is in a retrograde orbit with a projected angle of 123$^{\circ}$ between the orbital and the stellar rotation axes \citep{Queloz2010}. Hence, the prior assumption of an inclination angle of 90$^{\circ}$ is probably invalid,
and the photometric rotation period in combination with the v\,$\sin i$ indicate an inclination of the rotation axis with the line of sight of about 30$^{\circ}$.


\section{Observations and data analysis}
\label{Sect:obs}

\begin{figure*}
  \centering
  \includegraphics[width=0.32\hsize]{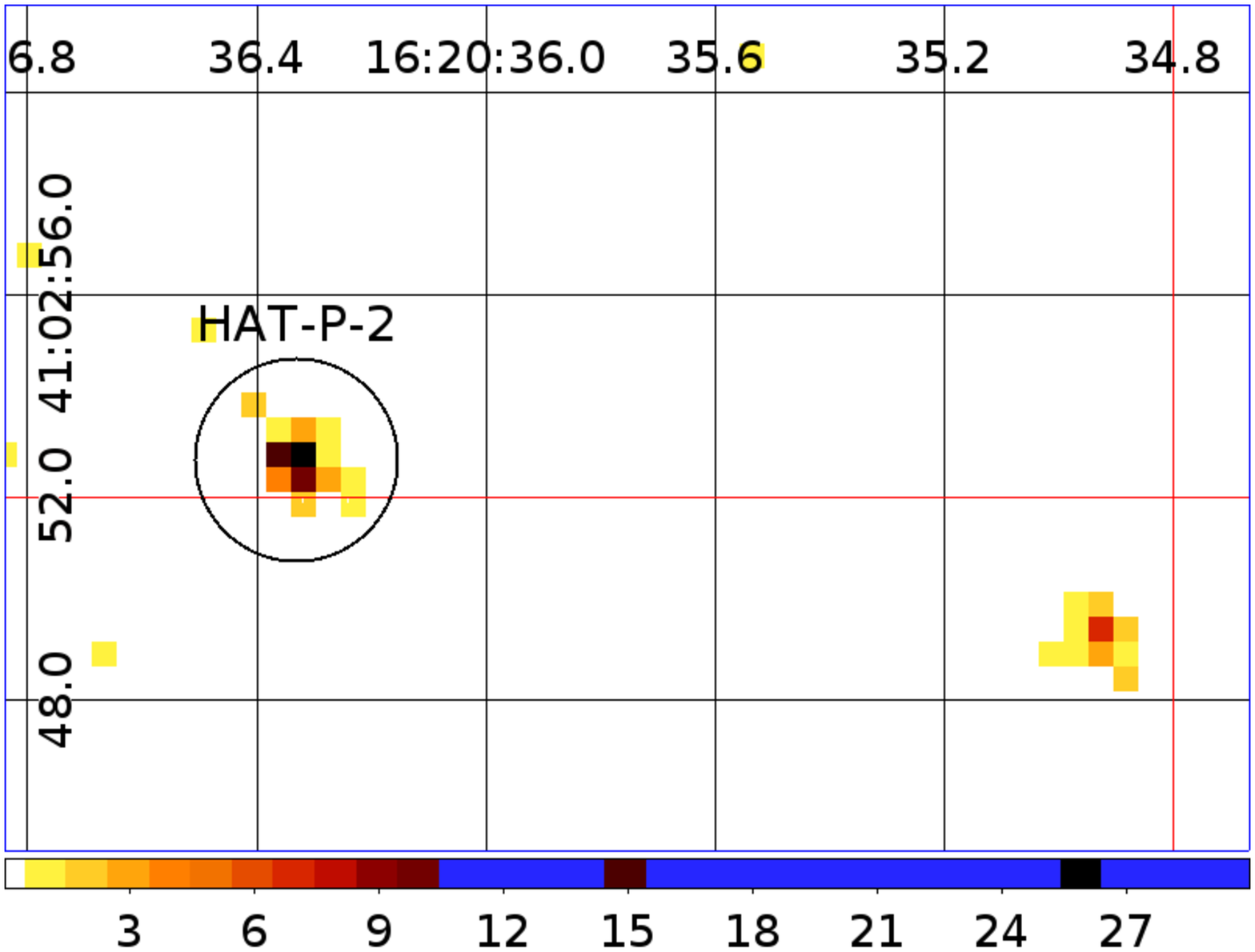}\hfill
  \includegraphics[width=0.32\hsize]{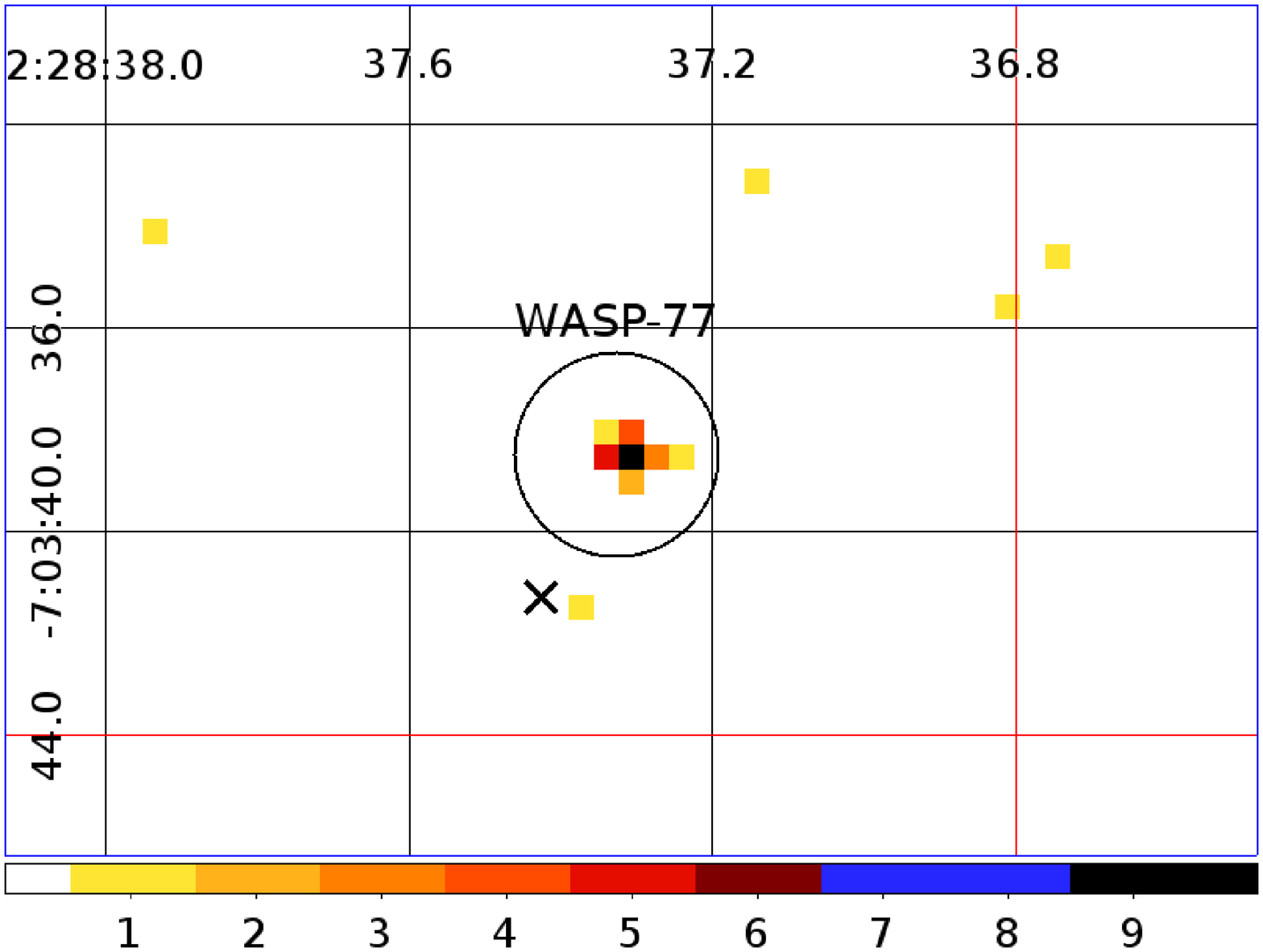}\hfill
  \includegraphics[width=0.32\hsize]{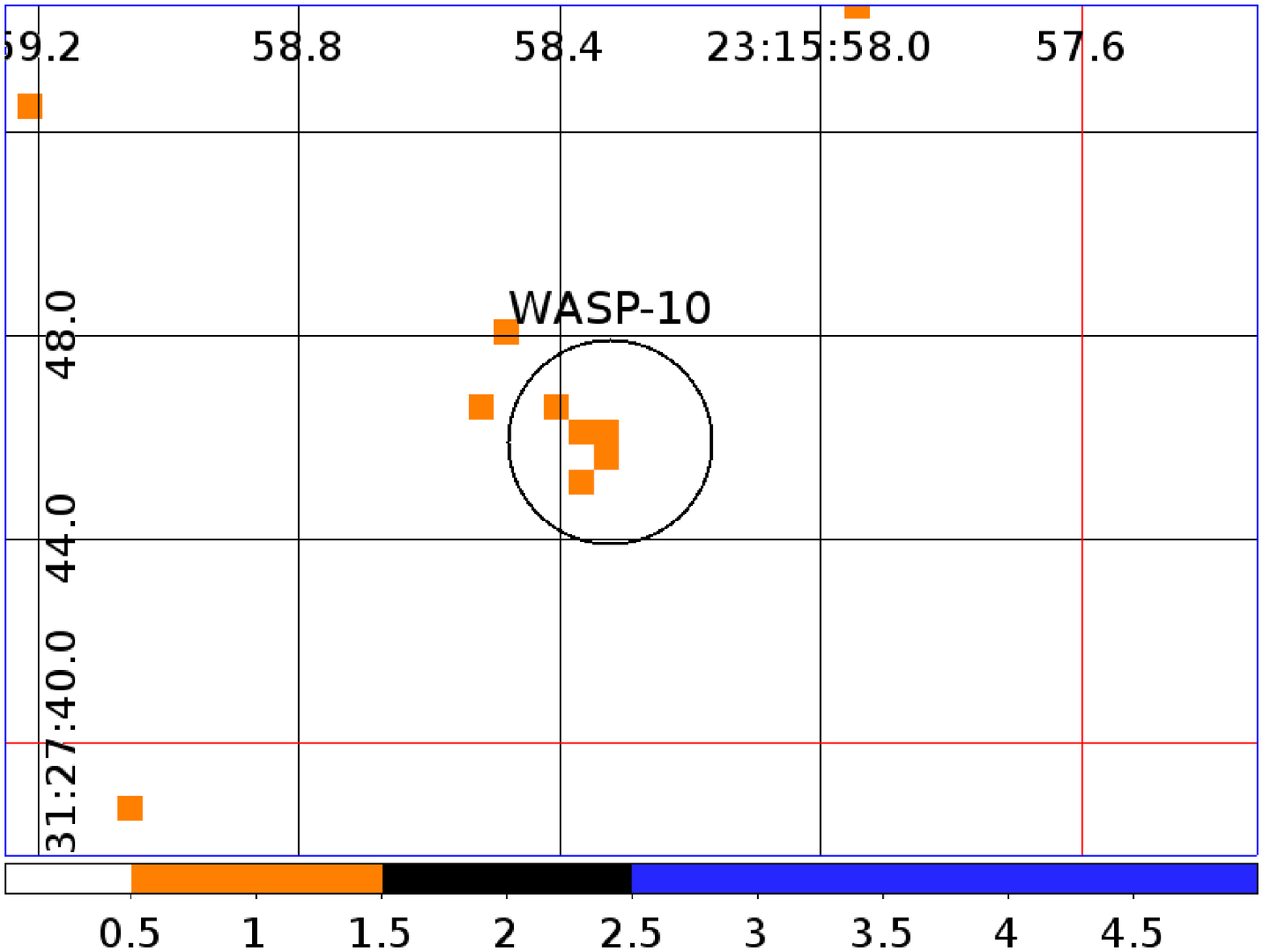}\\ \vspace{12pt}
  \includegraphics[width=0.32\hsize]{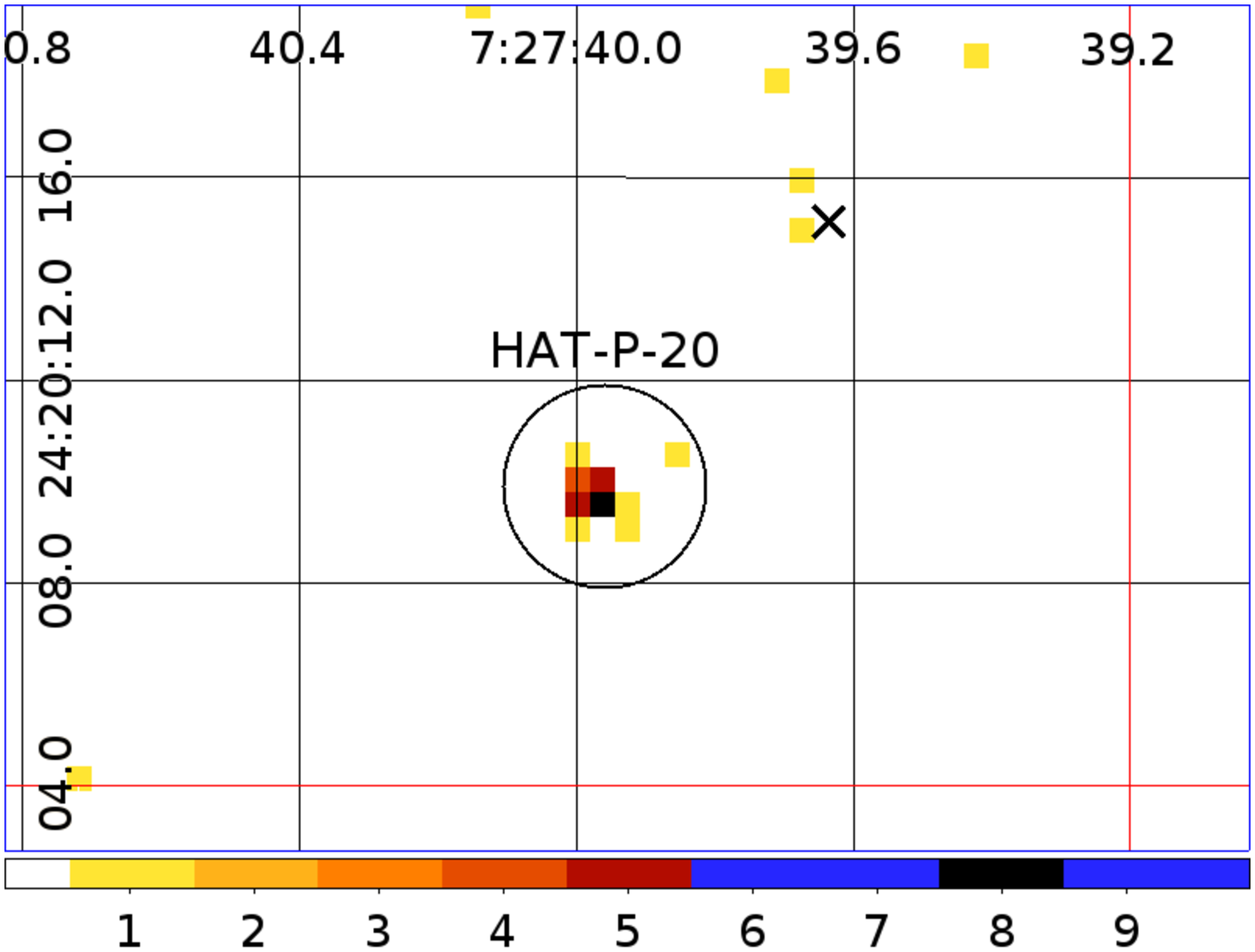}\hfill
  \includegraphics[width=0.32\hsize]{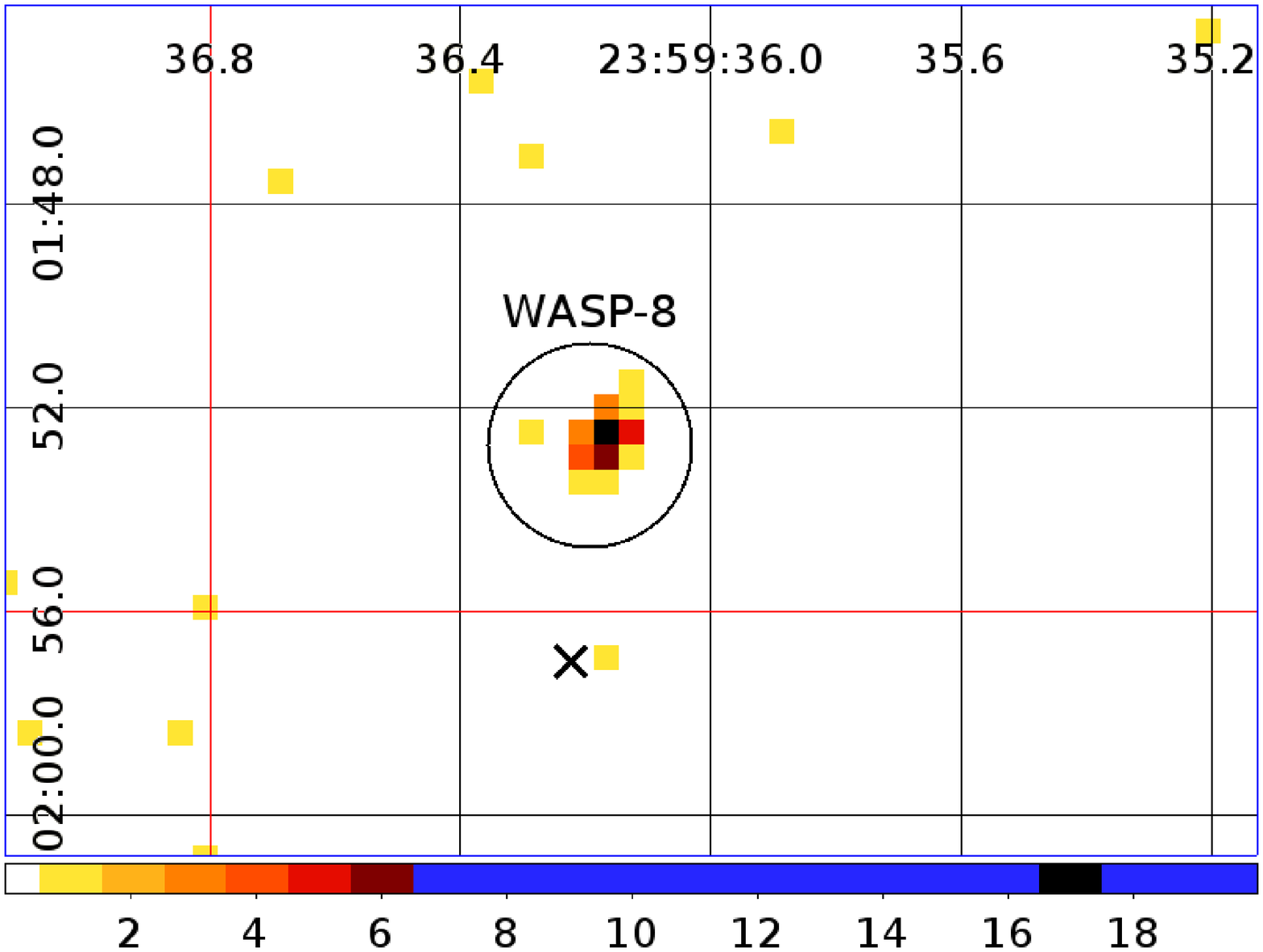}\hfill
  \includegraphics[width=0.32\hsize]{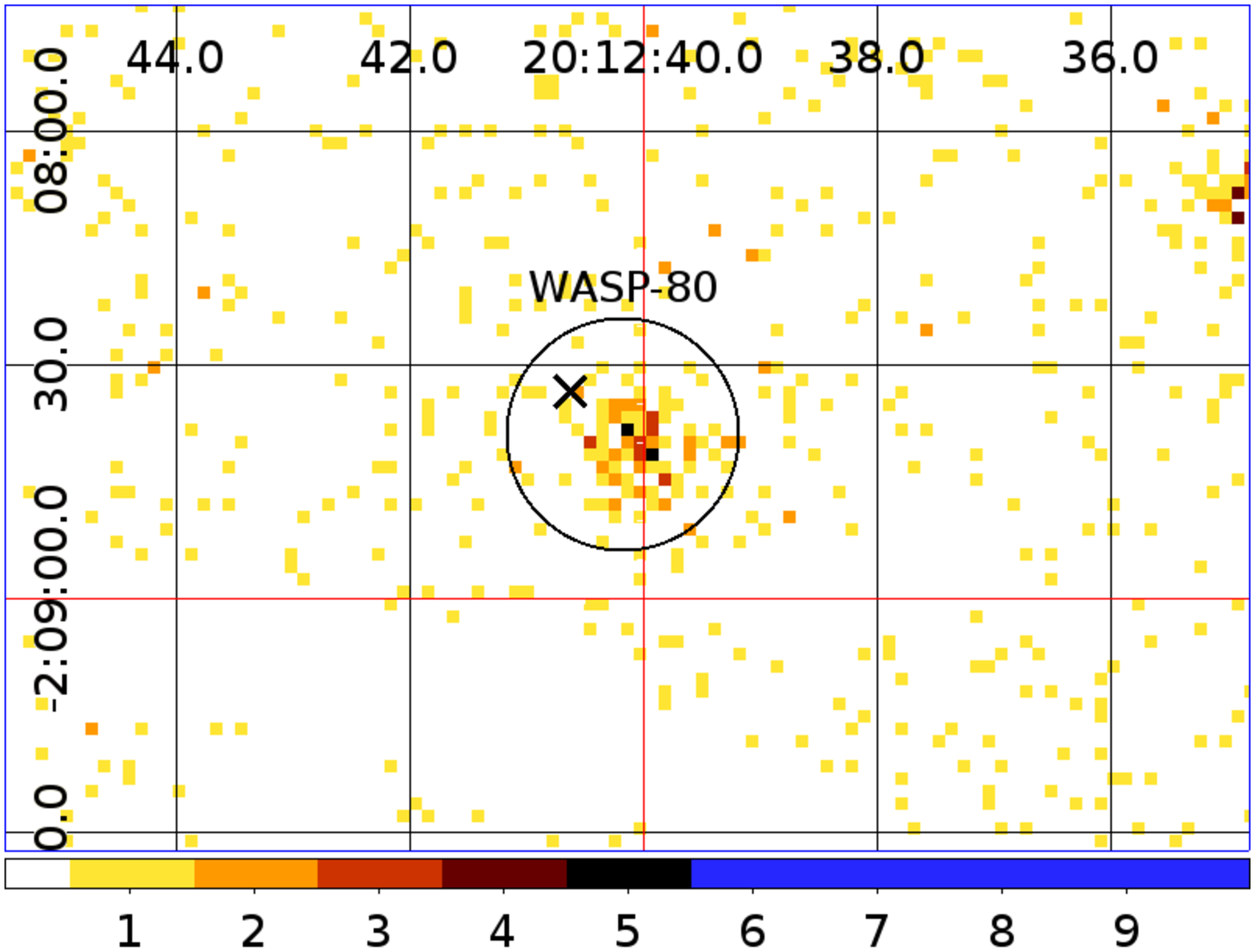}
  \caption{Source regions filtered for soft X-ray emission ($0.2-2.0$~keV).
           The target positions are marked by circles with a 2\arcsec{}/15\arcsec{} radius
           in the {\it Chandra}/{\it XMM-Newton} observations;
           the position of known companions is marked by crosses.
           In the panel of WASP-80, a nearby detected
           2MASS source is also marked by a cross.
           The X-ray source close to HAT-P-2 is discussed in
           Sect.~\ref{Sect:binaries}.
           Six out of seven targets have been detected as X-ray sources.
          }
  \label{FigXrayImgs}
\end{figure*}

We observed six planet hosts with the {\it Chandra} X-ray observatory using
10~ks long exposures and WASP-80 with {\it XMM-Newton} for 16~ks
(see Table~\ref{tab:obs}).
Three {\it Chandra} observations partially covered the planetary transit; the exposure of WASP-38 occurred completely within the transit of the hot Jupiter. 

We analyzed the {\it Chandra} ACIS-S observations  with the Chandra Interactive Analysis of Observations software package (CIAO) 4.6, CALDB 4.5.9 \citep{Fruscione2006}. We extracted the source counts in a circular region with a radius of 2\arcsec{} and background counts in another circular region with a 25\arcsec{} radius placed in a source-free region close to the target position. The target coordinates at the observational epoch were obtained from the SIMBAD database, accounting for proper motion. For HAT-P-20, we used the proper motion from the UCAC4 catalog \citep{Zacharias2013}.

In Fig.~\ref{FigXrayImgs} we show the source regions filtered for photon energies between $0.2-2$~keV, with the extraction regions marked by circles.
We detected six host stars  as X-ray sources (including the {\it XMM-Newton} observation, see below), only WASP-38 is a nondetection with zero photons at the source position.
The centroids of the source counts in the {\it Chandra} pointings show offsets smaller than 0.5\arcsec{} from the nominal source positions, while the Proposers' Observatory Guide (Version 16) cites a one sigma pointing error of 0.6\arcsec\footnote{\url{http://cxc.harvard.edu/proposer/POG/}}.
In the WASP-10 exposure, two photons were detected close to the source position, which shift the centroid by 1.1\arcsec{} if attributed to the target,
but only 3\% of the {\it Chandra} observations show a pointing error in excess of 1\arcsec{}.
The proper motion correction amounts to only 0.5\arcsec{} with a cited error of 10\%.
Accordingly, it is most likely that these are background photons, so they are not included in the analysis of WASP-10.
In any case, the two photons do not have a large impact on the analysis.

\begin{table}
  \caption{Optical Monitor results of WASP-80}
  \label{TabOM}
  \centering
  \begin{tabular}{l c c c}
    \hline\hline
    Filter & Cen. Wavelength & Flux\tablefootmark{a} & Mag.   \\
           & (nm)            &   & \\
    \hline
    U      & 344      & $5.4\pm0.5$ & \hphantom{$<$}\,14.5 \\
    UVW1   & 291      & \hphantom{$<$}\,1.5\hphantom{4} & \hphantom{$<$}\,16.0 \\
    UVM2   & 231      & \hphantom{$<$}\,0.1\hphantom{4}         & \hphantom{$<$}\,19.2 \\
    UVW2   & 212      & $<$\,0.04        &            $<$\,20.3  \\
    \hline
  \end{tabular}
  \tablefoot{The UVW2 filter did not yield a detection of the target.
             Count rate conversion for a K0V star ({\it XMM-Newton} User Handbook).
             The UVW1 and UVM2 conversion introduces a factor of two
             error and for UVW2 a factor of ten.
             Magnitudes were computed according to the fluxes.
             \tablefoottext{a}{($10^{-15}$~erg\,cm$^{-2}$s$^{-1}$\AA{}$^{-1}$)}
            }
\end{table}

WASP-80 was observed on May 15th, 2014, with all three detectors of the European Photon Imaging Camera (EPIC) on board of {\it XMM-Newton} in full frame mode with medium filters;
more details about the telescope and the detectors can be found in the {\it XMM-Newton} User Handbook\footnote{\url{http://xmm.esac.esa.int}}.
We also obtained four full frame
images with the Optical Monitor (OM), each 1320~s long with the filters U, UVW1, UVM2, and UVW2. The derived fluxes and magnitudes are given in Table~\ref{TabOM}.
We reduced the data  with the Scientific Analysis System (SAS 13.0.0) with the standard procedure and filters. The X-ray exposures are free of high background periods, so that no time filtering was applied. The source counts were extracted in a circular region with a radius of 15\arcsec{} and background counts were extracted from three circular regions on the same CCD, which were chosen with at least 30\arcsec{} distance to any recognizable X-ray source. 

\subsection{Binaries}
\label{Sect:binaries}

WASP-8, WASP-77, and HAT-P-20 are known visual binary stars and the positions of the faint red companions (BD-07 436B, 2MASS~J07273963+2420171, WASP-8~B) are marked in Fig.~\ref{FigXrayImgs} by crosses. The relative position of the companions was obtained from images of the Two Micron All Sky Survey \citep[2MASS,][]{Skrutskie2006} and transfered to the X-ray images, assuming a common proper motion of the binaries.
For the companion of HAT-P-20, the infrared colors J = 10.14, H = 9.44, and K = 9.22 from 2MASS indicate an M-type dwarf. The companions of WASP-77 and WASP-8 have spectral types of K5V and M, respectively \citep{Maxted2013,Queloz2010}.

The 2MASS images of all targets were screened for further close companions. Only
WASP-80 shows a 2MASS source (2MASS~20124062-0208333) about 4 magnitudes darker at 9\arcsec{} distance, which is also marked in  Fig.~\ref{FigXrayImgs}; the source is also clearly distinguishable in the OM exposure. The infrared colors indicate a late K to early M-type star, similar to the spectral type of WASP-80 \citep{Triaud2013}, which is not known to be a binary star. The magnitude difference at a similar spectral type suggests that the 2MASS detection is a background source.

In the exposure of HAT-P-2, we detected a soft X-ray source  at a distance of 16\arcsec{} from the target, corresponding to 1800~AU at the distance of the HAT-P-2 \citep[114~pc,][]{Hipparcos2007}. 
The 2MASS J-band yields no detection with an upper limit of 15.8~mag\footnote{\href{http://www.ipac.caltech.edu/2mass/overview/about2mass.html}{http://www.ipac.caltech.edu/2mass/overview/about2mass.html}},
hence a dwarf with a spectral type later than M5 would not be detected.
Assuming the distance of HAT-P-2, the X-ray source has a luminosity of $2\times10^{28}$~erg\,s$^{-1}$, whereas the maximal $L_{\text{X}}/L_{\text{bol}} = 10^{-3}$ in M-dwarfs results in $4\times10^{27}$~erg\,s$^{-1}$ \citep{James2000} . The light curve excludes a strong flare and the gyrochronological age of HAT-P-2 \citep[0.5~Ga in our analysis, 1.6~Ga in the more detailed analysis of][]{Brown2014} excludes increased activity due to a young age \citep[$<$\,100~Ma,][]{Stelzer2001}. 
We conclude that this source is most likely a background object, possibly of extragalactic origin because HAT-P-2 does not lie in the galactic plane.


\subsection{Light curve analysis}

\begin{figure}
  \centering
  \includegraphics[width=\hsize]{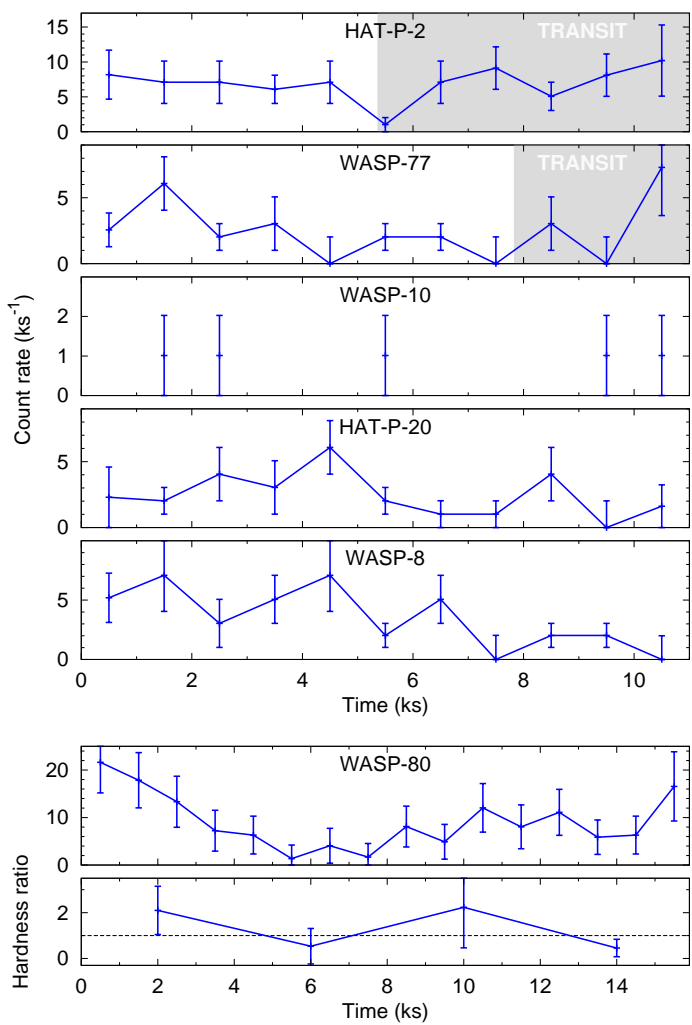}
  \caption{X-ray light curves of the targets with a 1~ks binning.
           The {\it XMM-Newton} observation of WASP-80 is about 5~ks longer,
           and the count rate is the sum of 
           all EPIC detectors; the error is given by the pipeline.
           In the {\it Chandra} observations, error bars represent the
           1-sigma Poissonian uncertainty on the count rate.
           Planetary transits during observations are marked by shaded areas.
           The WASP-10 panel shows only the arrival times of the five source
           photons.
           The last panel depicts the hardness ratio ($0.65-2.0 / 0.2-0.65$~keV) 
           in the WASP-80 observation.
           No pronounced flare occurred during the observations.
          }
  \label{FigXrayLC}
\end{figure}

For our analysis of the high-energy irradiation of planetary atmospheres, we are interested in the mean coronal emission of the targets. Strong flares during the observations would result in unreasonably high luminosity estimates.
The exposure time corrected X-ray light curves of the targets with 1~ks binning are shown in Fig.~\ref{FigXrayLC}. We added the background subtracted counts of the three EPIC instruments for the light curve of WASP-80. No background correction is applied in the {\it Chandra} exposures.
None of the observations shows a strong flare.

The count rate of WASP-80 declines significantly during the first 6~ks of the observation. The bottom panel of Fig.~\ref{FigXrayLC} shows the hardness ratio ($0.65-2.0 / 0.2-0.65$~keV) of the target. The ratio decreases along with the count rate, indicating that this trend can be related to prolonged flaring activity at the exposure start. However, the count rate around 6~ks is unusually low compared to the rest of the observation, so eventually we use the complete exposure to derive the mean X-ray flux of the target.

The observations of HAT-P-2 and WASP-77 partially cover the planetary transit. Individual X-ray exposures are not sensitive to the level of absorption expected from expanded atmospheres of hot Jupiters \citep[$\sim$\,7\%,][]{Poppenhaeger2013}.
Nevertheless, the periods affected by the planetary transits are shaded in Fig.~\ref{FigXrayLC}.
Comparing the count rate during the transit with the pretransit rate, we derive an upper limit for the size of an X-ray opaque planetary atmosphere assuming a homogeneous X-ray surface brightness of the host star.
For HAT-P-2 we have 31 in-transit counts with 32.7 counts expected, which corresponds to a 
maximum radius of an expanded atmosphere of 7.3~R$_{\mathrm{pl}}$, and for WASP-77 5/5.5 in-transit/scaled out of transit counts were observed resulting in upper limit of 6.3~R$_{\mathrm{pl}}$ for the planetary atmosphere (95\% confidence).

It is unlikely that the nondetection of WASP-38 is caused by an expanded planetary atmosphere that covers the complete stellar disk, because the atmosphere would have to be expanded over two stellar radii (24~$R_\text{pl}$), whereas usual estimates assume expansions by only a few planetary radii \citep[e.g., ][]{Vidal2003}. Although, the derived upper limit for the X-ray luminosity of the host star is low, the value is consistent with predictions (see Table~\ref{tabSim}) and with other activity indicators (see Sect.~\ref{SectOpt}).


\subsection{Spectral analysis and X-ray luminosities}

\begin{figure*}
  \centering
  \includegraphics[width=0.368\hsize, trim=0 1.3cm 0 0, clip]{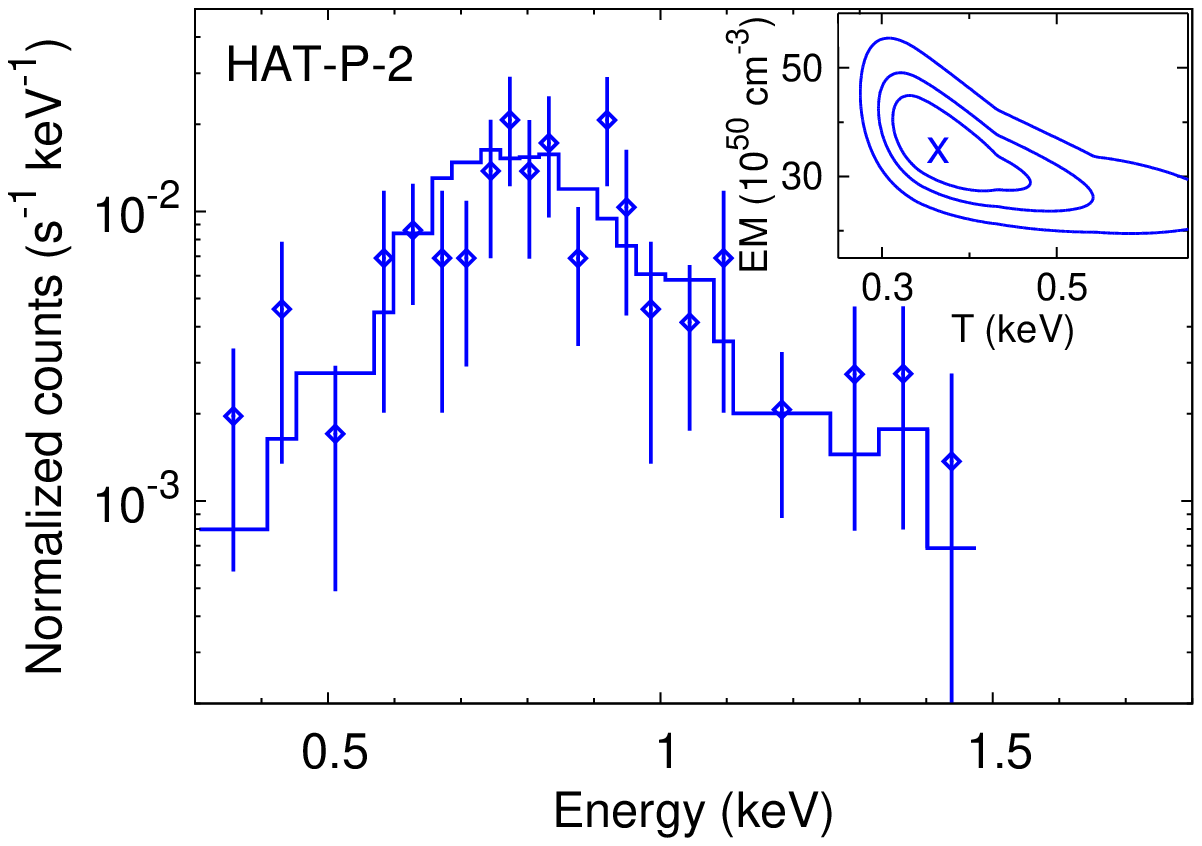}\hfill
  \includegraphics[width=0.31\hsize, trim=1.9cm 1.3cm 0 0, clip]{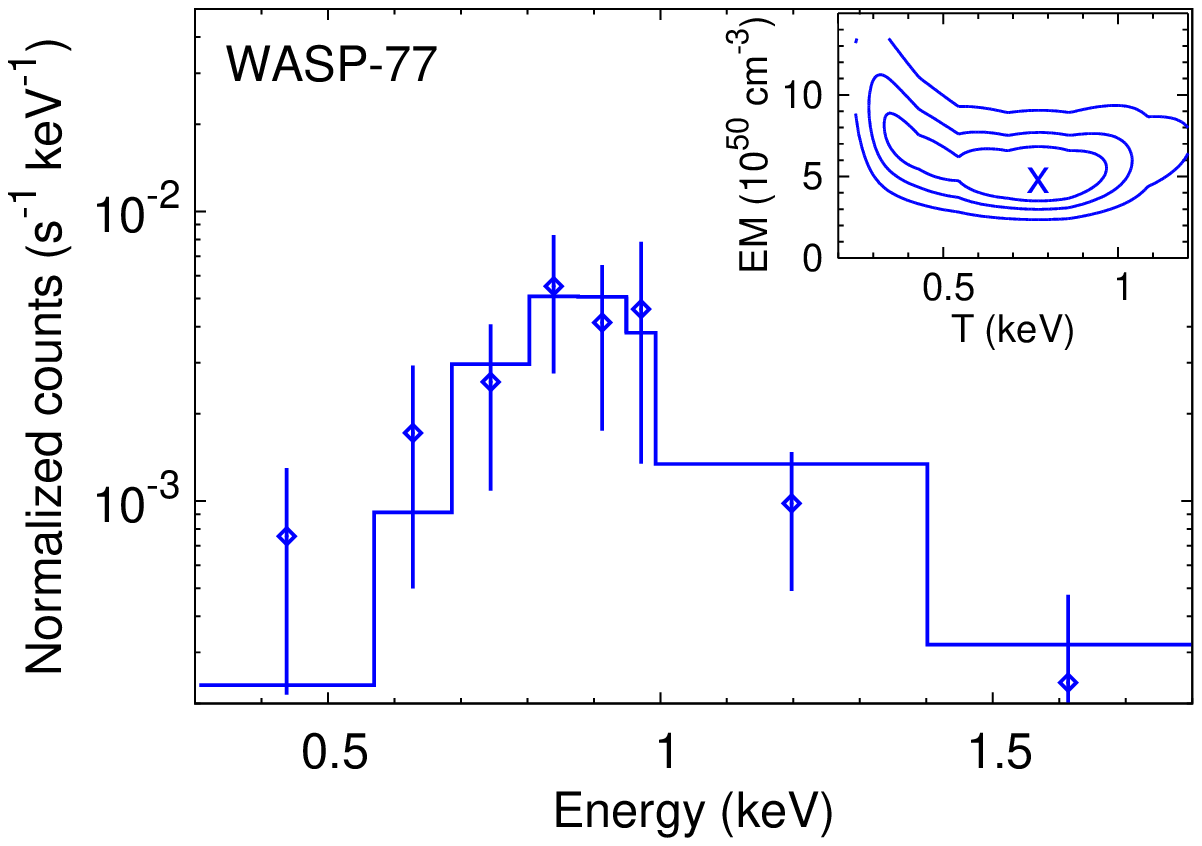}\hfill
  \includegraphics[width=0.31\hsize, trim=1.9cm 1.3cm 0 0, clip]{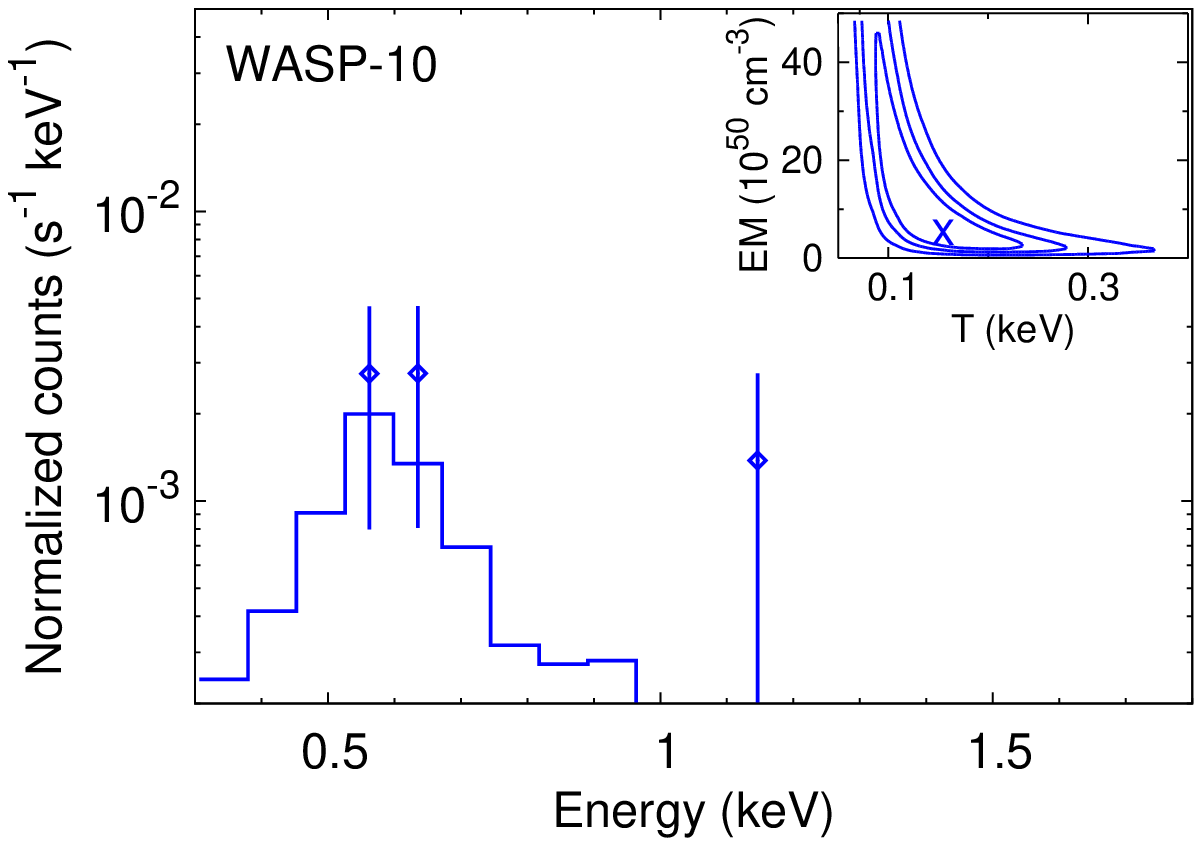}\\ \vspace{0.5pt}
  \includegraphics[width=0.368\hsize]{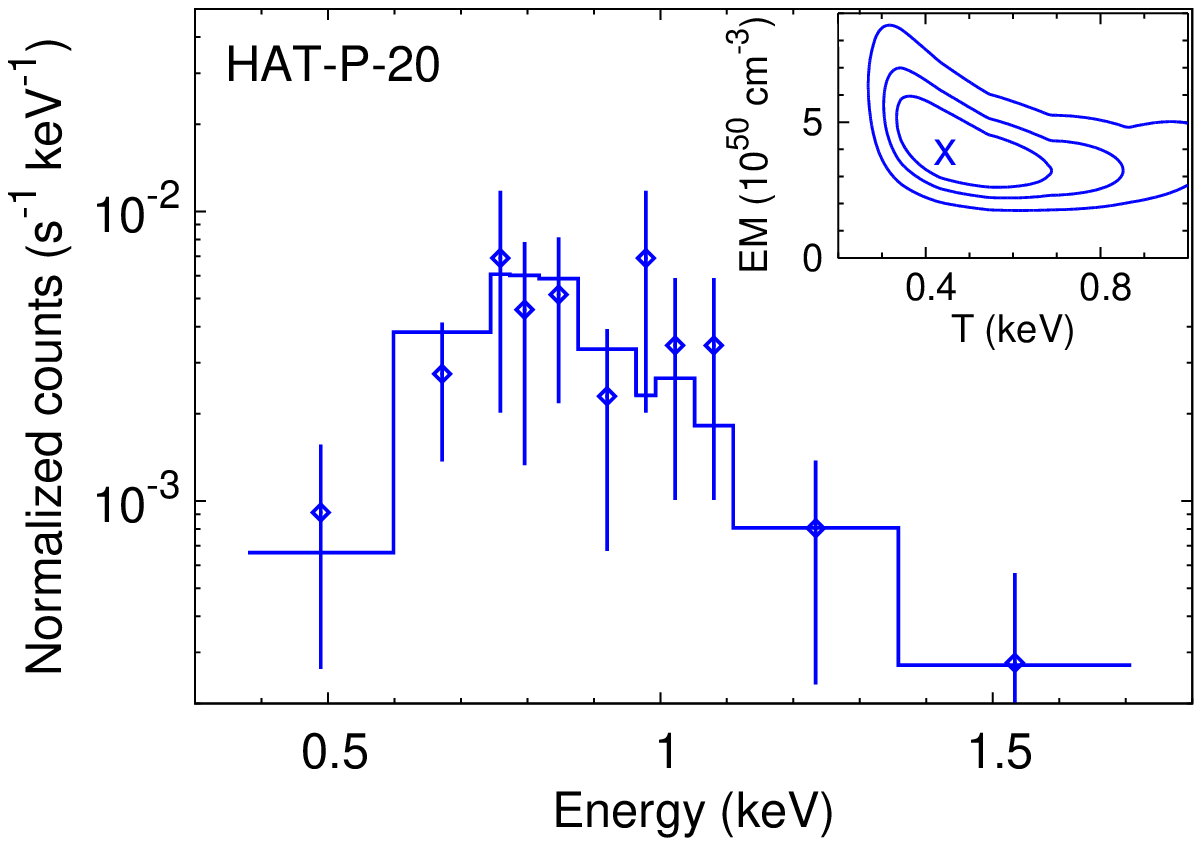}\hfill
  \includegraphics[width=0.31\hsize, trim=1.9cm 0 0 0, clip]{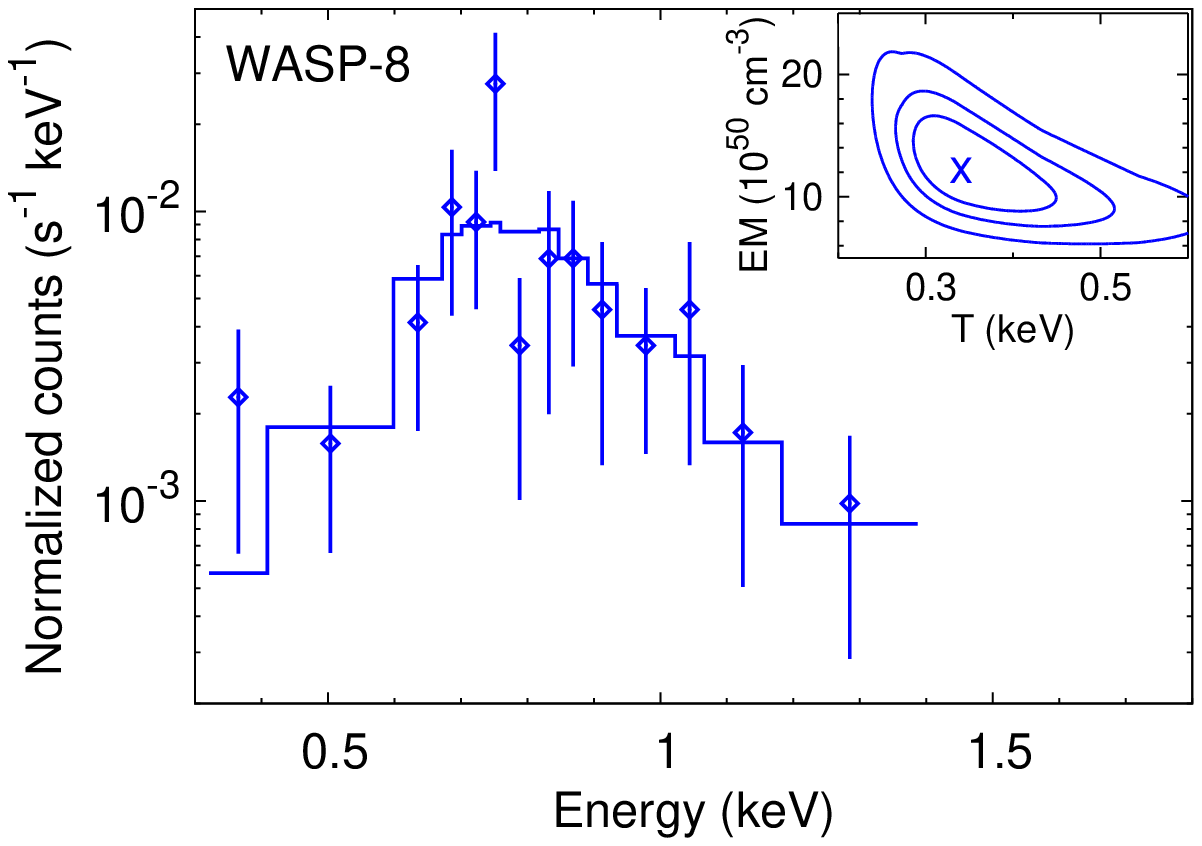}\hfill
  \includegraphics[width=0.31\hsize, trim=1.9cm 0 0 0, clip]{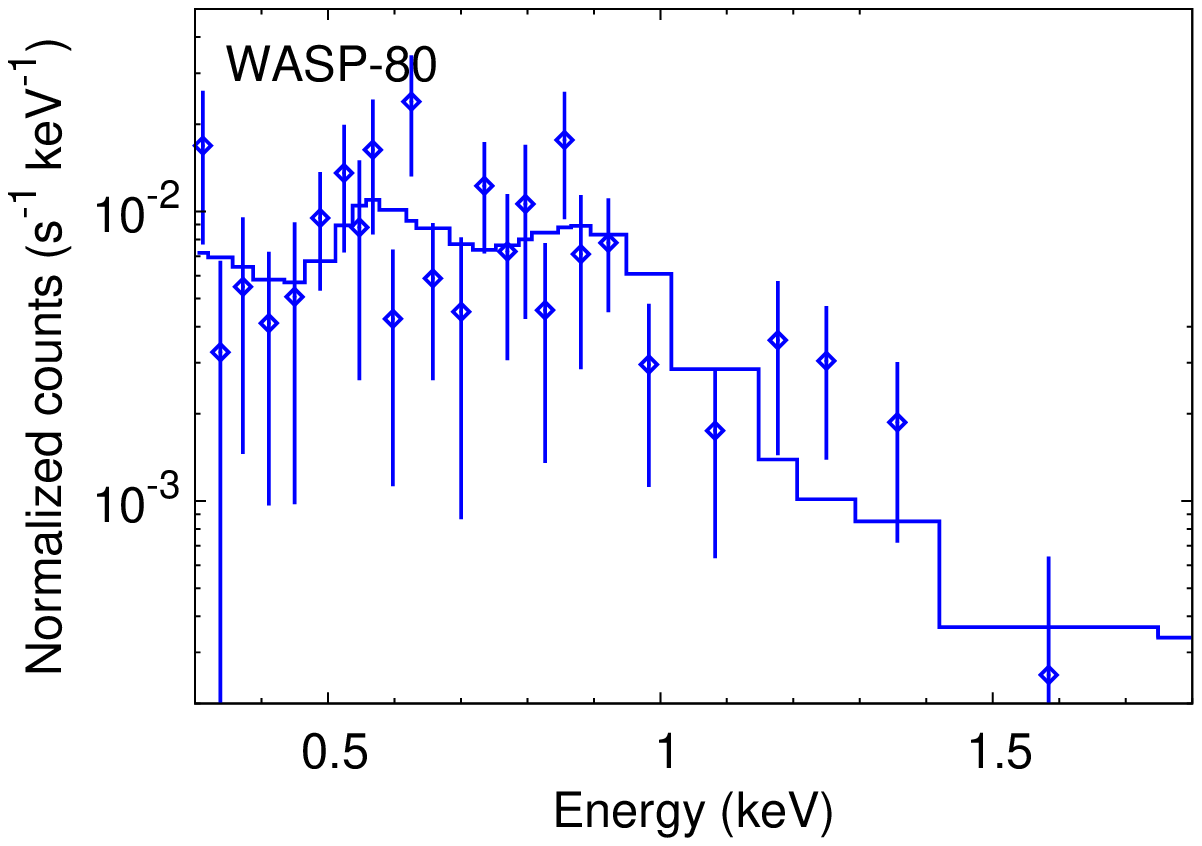}
  \caption{X-ray spectra of the targets.
           The plasma emission models are depicted by the histograms and the source count rates by diamonds with error bars.
           The $\chi^2$ contours for the 1, 2, and 3~$\sigma$ confidence intervals of the model parameters are shown in the inserts.
           WASP-80 was fitted with a two temperature model; the contours of the four parameters are not shown.
           With the exception of WASP-10, which only has five source counts, the model parameters are well confined.
          }
  \label{FigXraySpec}
\end{figure*}

Spectra were extracted from source and background regions
and analyzed with XSPEC V12.7.1 \citep{Arnaud1996}.
The three spectra from the individual cameras on board  {\it XMM-Newton} were fitted in a joint analysis.
We used the C-statistics, appropriate for low count rates \citep{Cash1979}. According to the XSPEC guidelines, the data were binned to contain at least one photon per bin\footnote{\url{http://heasarc.gsfc.nasa.gov/xanadu/xspec/}}
with the exception of WASP-10, for which the five source counts were fitted using the unbinned data.
We used  an absorbed, one temperature, optically thin plasma emission model \citep[APEC,][]{Foster2012} with solar abundances \citep{Grevesse1998}.  We added a second temperature component for WASP-80. In this case, the spectra contain sufficient source counts because of the longer exposure and a five times higher effective area at 1~keV of {\it XMM-Newton} compared with {\it Chandra}.

Since the absorption component of the model remained ill-constrained in the fitting procedure, we fixed the interstellar hydrogen column density using the source distance and an average interstellar hydrogen density of 0.1~cm$^{-3}$ \citep{Redfield2000}. The resulting spectra and models are shown in Fig.~\ref{FigXraySpec}, where the spectra were binned to a lower resolution for visualization.
The inserts show the 1,2, and 3 sigma confidence contours of the two model parameters: plasma temperature and emission measure.
These models are used to compute the X-ray fluxes in the 0.124 to 2.48~keV (5 to 100~\AA{}) spectral range (see Table~\ref{tabSim}).
The detected targets have mean coronal temperatures ranging from 2 to $9\times10^6$~K and their X-ray luminosities vary by one order of magnitude from 0.7 to $8\times10^{28}$~erg\,s$^{-1}$.

The short X-ray exposures do not yield detections of the faint companions in the three binary systems. However, one or two photons are detected at the expected positions of the secondaries. Assuming similar coronal parameters for both components, we derive upper limits for the X-ray luminosity of the B components (95\% confidence, see Table~\ref{TabBin}).

\subsection{Optical observations}
\label{SectOpt}

To check for activity indicators in the optical,
we obtained quasi-simultaneous optical spectroscopy for WASP-38 and WASP-77. In particular, the spectra were taken with
the ``Heidelberg Extended Range Optical Spectrograph'' (HEROS), mounted at the 1.2~m ``Telescopio Internacional de Guanajuato, Rob\'otico-Espectrosc\'opico'' (TIGRE) at La Luz observatory in Mexico \citep{Schmitt2014}.
The spectra cover the $350-880$~nm range with only a small gap around $570$~nm at a resolution of about 20\,000.

We observed WASP-38  for 30~min, less than one hour after the X-ray exposure, and
we observed WASP-77  twice, 5~h before and 11~h after the X-ray observation with each exposure lasting 45~min.
Both the WASP-38 spectrum and the combined WASP-77 spectrum show a 
mean signal-to-noise ratio (SNR) of 25 in the blue channel, which is sufficient to derive the Mount Wilson S-index
\citep[$S{\text{MWO}}$, ][]{Wilson1978}. We further converted $S{\text{MWO}}$ into the chromospheric $\log R'_{\text{HK}}$ index, which
represents the ratio of emission in the \ion{Ca}{ii} H and K emission lines and the stellar bolometric luminosity \citep{Noyes1984, Rutten1984}.

For WASP-38, we
 obtained values of $S_{\text{MWO}} = 0.153 \pm 0.006$ and $\log R'_{\text{HK}} = -4.87 \pm 0.04$,
which are consistent with a basal chromospheric flux level and thus a low level of chromospheric activity \citep{Mittag2013}.
This finding is also consistent with the absence of detectable X-ray emission during our observation.
For WASP-77, we derived an S-index, $S_{\text{MWO}}$, of $0.338 \pm 0.011$, corresponding to a
$\log R'_{\text{HK}}$ value of $-4.57\pm 0.02$. This
indicates a moderate level of chromospheric activity clearly exceeding the basal level, which is, again, consistent
with the detection of X-ray emission. Between the consecutive nights, we found no detectable variability in the
H$\alpha$ line, covered by the red spectral channel where the SNR is higher ($\sim$\,40).


\section{Planetary irradiation and mass loss}\label{Sect:results}

\begin{table*}
\setlength{\tabcolsep}{3.8pt}
\small
\caption{Results from the analysis of the X-ray observations and the mass loss analysis for the host stars in the sample.}             
\label{tabSim}      
\centering
\begin{tabular}{l @{\hspace{6pt}}l ccccc l cccccccc}
\hline\hline\vspace{-5pt}\\
    &
    &
  \multicolumn{5}{c}{X-ray analysis} & 
    &
  \multicolumn{8}{c}{Mass-loss analysis} \\
  \vspace{-7pt}\\ \cline{3-7}\cline{9-16} \vspace{-5pt}\\
  System &  
    &
  T & 
  EM\tablefootmark{A} & 
  $F_{\mathrm{X}}$\tablefootmark{B} & 
  $L_{\mathrm{X}}$ & 
  $L_{\mathrm{X}}^{\mathrm{rot}}$ & 
   &
  $\log L_{\mathrm{EUV}}$  &
  $\log L_{\mathrm{Ly}\alpha}$  &
  $F_{\mathrm{Ly}\alpha}$\!\tablefootmark{C} & 
  $F^{\mathrm{abs}}_{\mathrm{Ly}\alpha}$\!\tablefootmark{D} & 
  $F_{\mathrm{XUV}}$\!\tablefootmark{E} & 
  $\log \dot{M}$ & 
  $\log \dot{M}^{\mathrm{rot}}$ & 
  $\Delta M$ \\ 
  \vspace{-9pt}\\
    & 
    &
  ($10^6$\,K) & 
   & 
   &  
  \multicolumn{2}{c}{($10^{28}$~erg\,s$^{-1}$)} &
    &
  (erg\,s$^{-1}$) &
  (erg\,s$^{-1}$) & 
    &
    &
    &
  (g\,s$^{-1}$)  &
  (g\,s$^{-1}$) &
  (\%) \\
  \vspace{-7pt}\\ \hline\vspace{-5pt}\\ 
  HAT-P-2      && $4.2^{+1.4}_{-0.6}$  &            $35.8^{+10.3}_{-8.8}$                  & \hphantom{$<$}\,$5.2^{+0.5}_{-0.5}$  & \hphantom{$<$}\,$8.2^{+1.7}_{-1.5}$  & $39.$\hphantom{33} && \hphantom{$<$}\,28.9 & \hphantom{$<$}\,29.0 & \hphantom{$<$}\,6.5 & \hphantom{$<$}\,0.0\,$-$\,\hphantom{1}7.8 & \hphantom{$<$}\,1.3\hphantom{4} &  \hphantom{$<$}\,9.52 & \hphantom{1}9.79 & 0.004  \\ 
  \vspace{-6pt}\\
  WASP-38      &&  ---                 & ---                                               & $<$\,0.8\hphantom{$^{+0.3}$}         & $<$\,1.1\hphantom{$^{+0.3}$}         & \hphantom{3}$9.4$\hphantom{3} &&            $<$\,28.5 &            $<$\,28.7 &            $<$\,3.7 &            $<$\,0.3\,$-$\,\hphantom{1}0.4 &            $<$\,0.29            &             $<$\,9.42 & \hphantom{1}9.74 & 0.008  \\
  \vspace{-6pt}\\
  WASP-77      && $9.0^{+2.4}_{-5.9}$  & \hphantom{3}$5.0^{+2.0}_{-1.6}$\hphantom{$^{0}$}  & \hphantom{$<$}\,$1.3^{+0.3}_{-0.2}$  & \hphantom{$<$}\,$1.4^{+0.3}_{-0.3}$  & \hphantom{3}$2.1$\hphantom{3} && \hphantom{$<$}\,28.6 & \hphantom{$<$}\,28.8 & \hphantom{$<$}\,5.6 & \hphantom{$<$}\,1.3\,$-$\,\hphantom{1}4.9 & \hphantom{$<$}\,3.3\hphantom{4} &             \,\,10.85 & 10.3\hphantom{1} & 0.29\hphantom{1}  \\
  \vspace{-6pt}\\
  WASP-10      && $1.8^{+1.0}_{-1.8}$  & \hphantom{3}$6.0^{+79.0}_{-6.0}$                  & \hphantom{$<$}\,$1.3^{+0.1}_{-1.3}$  & \hphantom{$<$}\,$1.2^{+0.6}_{-1.2}$  & \hphantom{3}$1.8$\hphantom{3} && \hphantom{$<$}\,28.6 & \hphantom{$<$}\,28.7 & \hphantom{$<$}\,5.7 & \hphantom{$<$}\,0.0\,$-$\,\hphantom{1}6.2 & \hphantom{$<$}\,1.2\hphantom{4} &  \hphantom{$<$}\,9.90 & \hphantom{1}9.47 & 0.01\hphantom{1}  \\
  \vspace{-6pt}\\
  HAT-P-20     && $5.2^{+3.6}_{-1.4}$  & \hphantom{3}$4.0^{+2.1}_{-1.5}$\hphantom{$^{0}$}  & \hphantom{$<$}\,$1.7^{+0.2}_{-0.4}$  & \hphantom{$<$}\,$1.0^{+0.2}_{-0.2}$  & \hphantom{3}$1.1$\hphantom{3} && \hphantom{$<$}\,28.5 & \hphantom{$<$}\,28.7 & \hphantom{$<$}\,8.9 & \hphantom{$<$}\,1.9\,$-$\,14.6            & \hphantom{$<$}\,1.2\hphantom{4} &  \hphantom{$<$}\,9.22 & \hphantom{1}8.75 & 0.001  \\
  \vspace{-6pt}\\
  WASP-8       && $3.9^{+1.4}_{-0.7}$  & $12.5^{+4.5}_{-3.9}$\hphantom{$^{0}$}             & \hphantom{$<$}\,$3.1^{+0.4}_{-0.6}$  & \hphantom{$<$}\,$2.8^{+0.6}_{-0.7}$  & \hphantom{3}$1.9$\hphantom{3} && \hphantom{$<$}\,28.7 & \hphantom{$<$}\,28.9 & \hphantom{$<$}\,8.0 & \hphantom{$<$}\,3.5\,$-$\,13.3            & \hphantom{$<$}\,0.46            & \hphantom{$<$}\,9.53  & \hphantom{1}8.90 & 0.01\hphantom{1}  \\
  \vspace{-6pt}\\
  \multirow{2}{*}{WASP-80}  && $2.2^{+0.8}_{-0.9}$  & \hphantom{3}$1.7^{+ 2.3}_{-0.6}$\hphantom{$^{0}$}  & \multirow{2}{*}{\hphantom{$<$}\,$1.6^{+0.1}_{-0.2}$}  & \multirow{2}{*}{\hphantom{$<$}\,$0.7^{+0.5}_{-0.5}$}  & \multirow{2}{*}{\hphantom{3}$1.6$\hphantom{3}} && \multirow{2}{*}{\hphantom{$<$}\,28.5} & \multirow{2}{*}{\hphantom{$<$}\,28.7} & \multirow{2}{*}{\;10.8}  & \multirow{2}{*}{\hphantom{$<$}\,0.0\,$-$\,33.6} & \multirow{2}{*}{\hphantom{$<$}\,1.1\hphantom{4}} &             \multirow{2}{*}{\,\,10.54}  & \multirow{2}{*}{10.3\hphantom{1}} & \multirow{2}{*}{0.29\hphantom{1}}  \\
  \vspace{-8pt}\\
                            && $9.9^{+1.9}_{-1.9}$  & \hphantom{3}$1.2^{+ 0.4}_{-0.4}$\hphantom{$^{0}$}  &&&&&&&&&&&   \\
  \vspace{-3pt}\\
  WASP-43      && ---                  & ---                                               & ---                                  & \hphantom{$<$}\,$0.8^{+0.6}_{-0.3}$  & \hphantom{3}$0.76$ && \hphantom{$<$}\,28.5  & \hphantom{$<$}\,28.7 & \hphantom{$<$}\,6.2 & \hphantom{$<$}\,0.3\,$-$\,\hphantom{1}8.7 & \hphantom{$<$}\,6.6\hphantom{4} &            \,\,10.84 & 10.1\hphantom{1} & 0.21\hphantom{1}   \\
  \vspace{-6pt}\\
  WASP-18      && ---                  & ---                                               & ---                                  & $<$\,0.07\hphantom{$^{+0.3}$}        & $17.$\hphantom{33} &&            $<$\,28.0  &            $<$\,28.3 &            $<$\,1.8 &            $<$\,1.5\,$-$\,\hphantom{1}2.9 &            $<$\,0.97            &            $<$\,9.36  & 10.3\hphantom{1} & 0.001   \\
  \vspace{-3pt}\\
  HD\,209458   && ---                  & ---                                               & ---                                  & $<$\,0.03\hphantom{$^{+0.3}$}        & \hphantom{3}$4.0$\hphantom{3} &&            $<$\,27.8  & \hphantom{$<$}\,28.6 &   \;15.\hphantom{0} & 22 &            $<$\,0.11                 &            $<$\,9.94  & 10.6\hphantom{1} & 0.04\hphantom{1}   \\
  \vspace{-6pt}\\
  HD\,189733   && ---                  & ---                                               & ---                                  & \hphantom{$<$}\,$1.5^{+0.5}_{-0.4}$  & \hphantom{3}$2.1$\hphantom{3} && \hphantom{$<$}\,28.6  & \hphantom{$<$}\,28.4 &   \;60.\hphantom{0} & 190 & \hphantom{$<$}\,2.1\hphantom{4}     &            \,\,10.73   & 10.2\hphantom{1} & 0.28\hphantom{1}   \\
  \vspace{-6pt}\\
  55\,Cnc (b)  && ---                  & ---                                               & ---                                  & \hphantom{$<$}\,0.04\hphantom{$^{+0.3}$} & \hphantom{3}$0.31$ && \hphantom{$<$}\,27.7  & \hphantom{$<$}\,28.1 &   \;65.\hphantom{0} & 440 & \hphantom{$<$}\,0.01     &            ---         & --- & ---   \\
  \vspace{-6pt}\\
  GJ\,436      && ---                  & ---                                               & ---                                  & \hphantom{$<$}\,0.01\hphantom{$^{+0.3}$} & \hphantom{3}$0.06$ && \hphantom{$<$}\,27.1  & \hphantom{$<$}\,27.7 &   \;35.\hphantom{0} & 50 & \hphantom{$<$}\,0.06     &  \hphantom{$<$}\,8.92  & \hphantom{$<$}\,8.78 & 0.18\hphantom{1}   \\
  \vspace{-6pt}\\ \hline  
\end{tabular}
\tablefoot{Columns are:
           name of the system,
           coronal temperature and emission measure \tablefoottext{A}{\!($10^{50}$~cm$^{-3}$)}\!\!;
           X-ray flux at Earth \tablefoottext{B}{\!(0.124-2.48~keV, $10^{-14}$~erg\,cm$^{-2}$\,s$^{-1}$)}\!\!;
           X-ray luminosity and predicted luminosity \citep{Pizzolato2003};
           EUV luminosity \citep[100-912~\AA{},][]{Sanz2011}; 
           reconstructed Ly$\alpha$ luminosity\,\citep{Linsky2013};
           reconstructed Ly$\alpha$ flux at Earth \tablefoottext{C}{\!($10^{-14}$\,erg\,cm$^{-2}$\,s$^{-1}$)}\!\!;
           minimum and maximum Ly$\alpha$ flux at Earth after interstellar absorption \tablefoottext{D}{\!($10^{-15}$\,erg\,cm$^{-2}$\,s$^{-1}$)}\!\!;
           XUV flux at planetary distance \tablefoottext{E}{($<$\,912~\AA{},
           $10^{4}$~erg\,cm$^{-2}$\,s$^{-1}$)}\!\!;
           mass loss rate \citep{Sanz2011} and  rotation based estimate \citep{Ehrenreich2011};
           and the last column is the total fractional mass loss of the planet in its lifetime (see text).}
\tablebib{X-ray flux: WASP-43 \citep{Czesla2013}; WASP-18 \citep{Pillitteri2014}; HD\,209458, HD\,189733, 55\,Cnc, and GJ\,436 \citep{Sanz2011}.
           Reconstructed \lya{} flux: HD\,209458 \citep{Wood2005}, HD\,189733 \citep{Bourrier2013}, 55\,Cnc \citep{Ehrenreich2012}, and GJ\,436 \citep{France2013}.
           Measured \lya{} fluxes of HD\,209458, HD\,189733, and GJ\,436 are mean integrated out-of-transit fluxes in the
           original HST data.
}
\end{table*}

While EUV emission ($100<\lambda<912$~\AA{}) contributes about $\sim\,$90\% of the radiation power of hydrogen ionizing ($\lambda<912$~\AA{}) emission from inactive stars \citep{Sanz2011}, above 400~\AA{} it is completely absorbed by interstellar hydrogen.
To derive estimates for the planetary mass loss rate, which crucially depends on the radiative energy input, we have to assess the EUV emission of the host stars.

\subsection{Stellar EUV emission}
 
\begin{figure}
  \centering
  \includegraphics[width=1.\hsize]{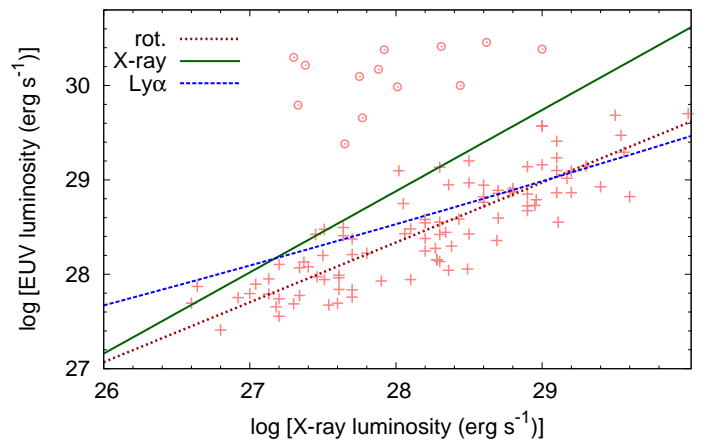}
  \caption{Estimates of stellar EUV emission in the 
           sample of \citet{Pizzolato2003} based on stellar
           rotation (red crosses), X-rays (green solid line) 
           and \lya{} emission (blue dashed line).
           A linear fit to the rotation based estimates is 
           represented by a red dotted line.
           Circles indicate stars with rotation periods
           shorter than 1~d. 
          }
  \label{FigCompEUV}
\end{figure}

Here, we compare three different methods to derive the EUV luminosity of dwarfs either based on the stellar rotation period, the X-ray luminosity, or the \lya{} luminosity.
The first method was introduced \citet{Lecavelier2007}, who apply a relationship between the stellar equatorial rotation velocity and the stellar flux in the S2 bandpass of the Wide Field Camera on board the {\it ROSAT} satellite \citep{Wood1994}. The flux in this band is further scaled to the full EUV range based on an average solar spectrum:
\begin{equation}\label{EqEUVlum_lecav}
  F_{\text{EUV}}\,(1\,\text{AU}) = 4.6 \left(\frac{\mathrm{v}_{\mathrm{eq}}}{2.0\,\text{km\,s}^{-1}} \right)^{1.4} ~\mathrm{erg\,cm}^{-2}\,\mathrm{s}^{-1} \,.
\end{equation}
Second, \citet[][]{Sanz2011} derive a relation between the X-ray (5 to 100~\AA{}) and EUV luminosities of main-sequence stars,
\begin{equation}\label{EqEUVlum}
  \log L_{\text{EUV}} = (4.80 \pm 1.99) + (0.860 \pm 0.073) \log L_{\text{X}} \,.
\end{equation}
Their method is based on a sample of stars with a full reconstruction of the emission measure distribution from 10$^{4}$ to 10$^{7}$~K.
The plasma model is then folded with an atomic emission model to derive the spectral energy distribution.
Third, \citet{Linsky2014} derive the EUV flux in several 100~\AA{} wide bands on the basis of the \lya{} flux. Their relations are based on intrinsic \lya{} fluxes, EUVE measurements ($<$\,400~\AA{}), and semiempirical models \citep[$400<\lambda<912$~\AA{},][]{Fontenla2014}.

To establish a comparability of these three approaches, we use a sample of main-sequence stars from \citet{Pizzolato2003}, which all have measured X-ray luminosities, known stellar rotation periods, and determined stellar masses.
To apply the relation from \citeauthor{Lecavelier2007}, we derive a stellar equatorial rotation velocity with the standard mass-radius relation for main-sequence stars \citep{Lacy1977} and then invert Eq.~\ref{EqRotation}.
The stellar \lya{} luminosity is needed for the \citeauthor{Linsky2014} method, but that luminosity is not available for the stars of Pizzolato.
However, a close correlation between the X-ray and \lya{} luminosities exists for main-sequence stars \citep{Linsky2013},
\begin{equation}\label{EqLyalum}
  \log L_{\text{lya}} = 19.7 + 0.322 \log L_{\text{X}} \,.
\end{equation}
The equation is a linear fit to the K5 to F5 stars in the sample from \citet{Linsky2013}. 
We can now use Eq.~\ref{EqLyalum} to derive the \lya{} luminosity, which is in turn utilized to compute the EUV flux. This method results in a second X-ray based estimate, and is valid for K5 to F5 stars. Among our sample, WASP-80, WASP-43, and especially GJ\,436 have a later spectral types resulting in a larger uncertainty of the derived EUV luminosity.

Figure~\ref{FigCompEUV} shows the three estimates for the EUV luminosities of the \citet{Pizzolato2003} sample versus the X-ray luminosities. The rotation based estimates show the expected scatter, which reflects the accuracy of the correlation between X-ray emission and rotation period. For a better visualization, we show a linear fit to the rotation based estimates, leaving out stars with periods shorter than 1~d because they are in the saturated regime.
Each method results in significantly different EUV luminosities. 
The best agreement occurs for inactive stars ($\log L_{\text{X}} = 27.2 $), where the three correlations agree within a factor of 2.5, but for highly active stars the estimate of \citeauthor{Sanz2011} is about one order of magnitude higher than both other relations.
The different results of the three approaches give an idea of the uncertainty in deriving stellar fluxes in this spectral range. The uncertainty is larger than the long-term variability expected for solar type stars. For example, in the Sun the chromospheric emission changes by up to 120\% over one solar cycle \citep{Woods2005}, which is a factor of two smaller than the best agreement between the EUV estimates.

For the following mass loss analysis we use the relation from \citet{Linsky2014} to derive the EUV luminosity of the host stars in our sample (see Table~\ref{tabSim}).
The relation is based on a detailed analysis of the EUV fluxes in eight spectral bands, and we consider the combination of observations and chromospheric models, validated on the basis of solar spectra, to be the best currently available option.


\subsection{Improved \lya{} flux estimates and interstellar absorption}\label{SectInterstABS}

Using Eq.~\ref{EqLyalum}, we derive improved estimates for the \lya{} luminosities of our targets.
The mean dispersion of this equation is a factor of two smaller than our initial estimates for the \lya{} luminosities based on the spectral type and the stellar rotation period (see Sect.~\ref{Sect:sample}).
We provide the new estimates in Table~\ref{tabSim} along with the expected unabsorbed stellar flux at Earth. These values can be directly compared with the unabsorbed flux of HD\,209458 \citep[$F_{\text{Ly}\alpha} = 15\times 10^{-14}$~erg\,cm$^{-2}$\,s$^{-1}$,][]{Wood2005}, where an expanded atmosphere has been unambiguously measured.
However, interstellar absorption reduces the observable emission line strength at Earth. The absorption depends on the interstellar neutral hydrogen column density, on the relative velocity of the star compared to the moving interstellar clouds, and on the shape of the \lya{} line. 

The structure of the interstellar medium has been determined within 15~pc around the Sun \citep{Redfield2008}, but our targets are found at distances $>$\,60~pc. Therefore, we use \ion{Na}{i} measurements of interstellar absorption of stars with similar lines of sight as our targets to estimate the neutral hydrogen column density.
From Table~1 of \citet{Welsh2010}, with \ion{Na}{i} absorption measurements along 1857 lines of sight,
we select stars within 8\deg{} angular distance of our targets resulting in three lines of sight for WASP-80, for example, and up to 16 lines of sight for HAT-P-2. 
The column densities are scaled linearly from the distances of the stars of \citeauthor{Welsh2010} to the distance of our targets and then converted into a hydrogen column density \citep{Ferlet1985}.

To evaluate the absorption, we constructed a mean \lya{} line profile using a double Gaussian combining a narrow (130~km\,s$^{-1}$) and a broad (400~km\,s$^{-1}$) emission line component, which typically describes the line profile well \citep{France2013}. The line flux is normalized to the values derived from Eq.~\ref{eqLyalpha}, with the broad component containing 15\% of the flux. For main-sequence stars, the \lya{} line profiles often show a double-peaked structure \citep{Wood2005}, which increases the detectable line flux compared to our estimates. The line width of stars also varies, imposing further uncertainty on the estimates (broader lines are less absorbed).
The standard \lya{} absorption cross section with a Voigt-profile is used for the absorption with a Doppler-width of 10~km\,s$^{-1}$. Deuterium absorption is included according to a Deuterium fraction in the Local Bubble of $1.56\times 10^{-5}$ \citep{Wood2004}.
The relative velocity of the absorber is neglected because \citet{Welsh2010} did not publish the kinematics of the interstellar medium. 
This introduces an uncertainty of a few 10\%, with absorbed fluxes up to 40\% higher for relative velocities of up to $\pm50$~km\,s$^{-1}$.

The absorbed line flux is computed for the minimum and maximum hydrogen column densities scaled from the surrounding measured lines of sight (see Table~\ref{tabSim}). With this procedure we take into account the uncertainty in the derived column densities due to inhomogeneous interstellar absorption, which is about an order of magnitude for most targets. For comparison, Table~\ref{tabSim} also contains the measured line fluxes of the four stars with detected absorbtion signals. The limiting flux for the Space Telescope Imaging Spectrograph on board the
Hubble Space Telescope is about $2.5\times 10^{-15}$~erg\,cm$^{-2}$\,s$^{-1}$ \citep[upper limit derived for GJ\,1214,][]{France2013}.  For most targets the estimates remain inconclusive.
However, the derived line fluxes for WASP-38 and WASP-18 are too low for a detection because of the inactive state of the stars; these systems are thus not well suited for \lya{} transit spectroscopy. Despite the large uncertainty in the absorbed \lya{} flux of WASP-80, this is the closest target among the sample and one of the most promising systems for follow-up campaigns.


\subsection{Mass loss analysis}\label{Sect:results2}

With the computed EUV fluxes, we can now determine the mass loss rates from the total radiative flux $F_{\text{XUV}} = F_{\text{X}} + F_{\text{EUV}}$, which impinges on the atmospheres \citep{Erkaev2007, Sanz2011}
\begin{equation}\label{EqMloss}
  \dot{M} = \frac{3\,\eta F_{\text{XUV}}}{4\,KG\,\rho_{\text{pl}}} \, .
\end{equation}
Here, $\rho_{\text{pl}}$ is the planetary density and $G$ denotes the gravitational constant. We adopt a heating efficiency of $\eta = 0.15$, which accounts for the fraction of the radiative energy needed for ionization processes or lost through radiative cooling.
This choice agrees with recent results from \citet{Shematovich2014}, who find $\eta < 0.20$ in the atmosphere of HD\,209458\,b.
Furthermore, it enables a direct comparison with the mass loss rates determined by \citet{Ehrenreich2011}, who utilized the stellar rotation based method from \citet{Lecavelier2007} to estimate the EUV fluxes (see Sect.~\ref{Sect:results}).
The parameter $K$ is a reduction factor for the gravitational potential of the planet due to tidal forces \citep{Erkaev2007}. It is given by
\begin{equation}\label{EqK_factor}
  K(\xi) = 1 - \frac{3}{2\xi} + \frac{1}{2\xi^3} \, ,
\end{equation}
with $\xi = r_{\mathrm{RL}}/r_{\mathrm{pl}} \approx (\delta/3)^{1/3} \lambda$, where $\delta = M_{\mathrm{pl}}/M_{\mathrm{st}}$ is planet to star mass ratio, and $\lambda = a/r_{\mathrm{pl}}$ the ratio of the semimajor axis to the planetary radius. 
All results are summarized in Table~\ref{tabSim}.

The X-ray based mass loss rates of our targets are on average higher than the rotation based estimates from \citeauthor{Ehrenreich2011}, which corresponds to the expected offset due to the different methods of deriving the EUV irradiation. However, based on our measured irradiation levels, we find the mass loss rate of WASP-38 to be more than a factor of two lower and the value of WASP-8 a factor of four higher than the previous estimates.
The mass loss rate of WASP-18 is even one order of magnitude lower than the stellar rotation based estimate.

We caution that these estimates are based on the assumption of energy-limited atmospheric escape, which in general provides an upper limit to the mass loss rate of a planet \citep{Watson1981}. 
Furthermore, for the heating efficiency, values ranging from 0.1 to 1.0 are used in literature \citep{Ehrenreich2011}, although the recent study of \citet{Shematovich2014} for HD\,209458\,b somewhat restricts this unconfined parameter.
This adds to a considerable source of uncertainty in the derivation of the high-energy irradiation (see Sect.~\ref{Sect:results}).
Hence, the mass loss rates must be viewed as order of magnitude estimates, but are more reliable than previous estimates without determination of the high-energy irradiation level \citep[e.g., ][]{Ehrenreich2011}.

The current fractional mass loss of all planets in the sample is small. 
The close proximity and the small planetary masses of WASP-77, WASP-43, and WASP-80 result in the strongest mass loss rates on a par with the mass loss rate of HD\,189733\,b.
WASP-80 loses 0.10\% of its mass in 1~Ga, assuming constant stellar emission.
Moreover, high-energy emission is up to a factor of 100 stronger in young stars (age\,$<100$~Ma) \citep{Ribas2005, Stelzer2001}. Assuming this high irradiation level for the first 100~Ma, the planet would have lost additional 0.28\% of its mass. 
For our targets, we derive the total mass loss estimate by combining this value and applying the present mass loss rate for the remaining lifetime (for the age estimates see Sect.~\ref{Sect:results3}).
According to these values, six hot Jupiters have lost less than 0.01\% of their masses through photoevaporation (see Table~\ref{tabSim}).
Even if we conservatively assume an uncertainty as large as a factor 100 for the total mass loss estimates,
these six planets cannot have lost more than 1\% of their mass because of photoevaporation during their lifetime. 

A final remark concerning photoevaporation: HAT-P-2 is the strongest X-ray source among our targets, but the mass loss rate is small because the planet is massive and has a comparatively large semimajor axis. 
Also the predicted mass loss rates of HAT-P-20\,b and WASP-18\,b are very small. These planets exhibit the highest densities in our sample. 
In general the formation of a hydrodynamic wind from such compact and massive objects is disputable. Hydrodynamic escape is only possible if the sonic point in the escape flow occurs before the exobase level is reached, otherwise radiative energy cannot be transformed into a bulk flow and only Jeans escape proceeds with a significantly lower mass loss rate \citep[e.g.,][]{Tian2005}.
The compactness of HAT-P-2\,b, HAT-P-20\,b, and WASP-18\,b may completely prevent the formation of a planetary wind.


\section{Angular momentum transfer from hot Jupiters to their host stars}\label{Sect:results3}

Giant gas planets in tight orbits raise substantial tidal bulges on their host stars \citep{Cuntz2000}. If the stellar rotation period is longer than the orbital period of the planet, it is possible that the tidal interaction induces a torque on the host star and reduces the stellar spin-down with age \citep{Poppenhaeger2014}. 
Based on measured X-ray luminosities \citeauthor{Poppenhaeger2014} predict stellar ages
of binaries, where the primary features a hot Jupiter in a close orbit.
Binaries have the advantage of providing two individual age estimates for the system, and a younger age of the primary would indicate continuous angular momentum transfer from the hot Jupiter's orbital motion.
Indeed, the two host stars with the strongest tidal interactions (HD\,189733, CoRoT-2) in the study of \citeauthor{Poppenhaeger2014}, suggest a significantly younger age of the hot Jupiter host compared with the secondary. In contrast three systems with smaller tidal interactions do not exhibit age differences. 

\begin{table}
  \small
  \caption{Age estimates for the binary systems within our sample}
  \label{TabBin}
  \centering
  \begin{tabular}{l@{\,\,\,}l c@{\hspace{6pt}}c@{\hspace{6pt}}c@{\hspace{6pt}}c@{\hspace{6pt}}c@{\hspace{6pt}}c@{\hspace{6pt}}c }
    \hline\hline\vspace{-5pt}\\
    System       & &Sp.T  & Sep.                 & $\log L_{\mathrm{X}}$   &  A$_{\text{iso}}$  &  A$_{\text{gyro}}$  & A$_{\text{X}}$ & $h_{\text{tide}}/H_{\text{P}}$ \\
                 & &      & (\arcsec{})          & (erg\,s$^{-1}$)         &  (Ga)                &  (Ga)                 & (Ga)                 &     \\
    \vspace{-7pt}\\\hline \vspace{-5pt}\\
    \multirow{2}{*}{WASP-77}  & A & G8V   & \multirow{2}{*}{3.3} & \hphantom{$<$}\,28.1 & \hphantom{$<$}\,5.3            & 1.7 &  \hphantom{$<$}\,4.5   & 0.12\hphantom{2} \\
                              & B & K5V   &                      & $<$\,27.5            &                                &       & $>$\,8.9               &  \\
    \vspace{-5pt}\\
    \multirow{2}{*}{HAT-P-20} & A & K3V   & \multirow{2}{*}{6.9} & \hphantom{$<$}\,28.0 & \hphantom{$<$}\,7\hphantom{.3} & 0.8   &  \hphantom{$<$}\,1.9   & 0.066  \\
                              & B & M     &                      & $<$\,27.4            &                                &       & $>$\,1.7               &   \\
    \vspace{-5pt}\\
    \multirow{2}{*}{WASP-8}   & A & G8V   & \multirow{2}{*}{4.8} & \hphantom{$<$}\,28.4 &           $<$\,3.6             & 1.6   &  \hphantom{$<$}\,1.6   & 0.002 \\
                              & B & M     &                      & $<$\,27.6            &                                &       & $>$\,1.5               &  \\
    \vspace{-7pt}\\\hline
  \end{tabular}
  \tablefoot{Columns are: system name and component, spectral type, separation of the 
             components, X-ray luminosity, isochrone age \citet{Brown2014, Bakos2011},
             gyrochronological age,
             age estimate based on the X-ray luminosity \citep{Poppenhaeger2014}, 
             and height of the tidal bulge in reference to the photospheric scale height.
            }
\end{table}

The angular momentum transfer was also studied by \citet{Brown2014}, who analyzes the gyrochronological and isochrone ages of exoplanet hosts. Reduced spin-down of the host stars due to tidal interactions would appear as young gyrochronological ages compared with the isochrone ages in hot Jupiter bearing systems.
\citeauthor{Brown2014} found indications for a general age difference of 1.8~Ga between gyrochronological and isochrone age estimates, but no correlation with tidal interactions. However, the author does not exclude effects in individual systems. 
We use the mean of the three color based rotation-age relations from \citet{Brown2014}\footnote{For the two F-type stars we use the original relation from \citet{Barnes2007} for the age relation based on $(B-V)$.} to derive gyrochronological age estimates for our targets (see Table~\ref{tabSysPara}).

For the three binaries among our targets we derive X-ray based age estimates, using the relationship from \citet{Poppenhaeger2014} .
The upper limits for the X-ray luminosity of the secondaries provide lower limits for their ages.
The strength of the tidal interactions can be assessed by the fractional height of the tidal bulges in reference to the photospheric pressure scale height $h_{\text{tide}}/H_{\text{p}}$ \citep{Cuntz2000}.
In addition, we provide the isochrone ages of the three targets from literature in Table~\ref{TabBin}.

For WASP-77\,A, we derive an X-ray age of about 5~Ga and the B component should be older than 9~Ga. For HAT-P-20\,A and B as well as for WASP-8\,A and B, we derive ages of about 2~Ga with only slight differences, but the upper limits indicate that the B components seem to be about twice the age of the primaries.
In contrast to WASP-8, WASP-77, and HAT-P-20 exhibit low X-ray luminosities compared with the rotation-based estimates. Their gyrochronological age estimates are consistently smaller than the X-ray based ages.
The hot Jupiter WASP-77\,b raises the highest relative tidal bulge on its host star, and this is the only system with an apparent age difference between the host star and the companion.
The isochrone age of this system is consistent with the X-ray age of the secondary, which further supports an age difference between the two stellar companions. The age difference also exceeds the general offset of the isochrone age estimates of 1.8~Ga found by \citeauthor{Brown2014}.
The same argument holds for HAT-P-20, but here the upper limit for the X-ray luminosity of the secondary does not reveal a significant age difference. There is no difference in the four age estimates of
WASP-8, which exhibits the weakest tidal interactions.
At this point the data do not provide strong evidence for a transfer of angular momentum from the hot Jupiters to the host stars.

\section{Conclusion}
\label{Sect:conclusion}

We measured the X-ray luminosities of seven hot Jupiter hosts and determined the level of high-energy irradiation and the planetary mass loss rates. Additionally, the two previously analyzed targets and four systems with detected atmospheres are included in our discussion.
According to our estimates, six of the eleven planets did not lose more than 1\% of their mass as the result of a hydrodynamic planetary wind during their lifetime. In our sample, WASP-80\,b, WASP-77\,b, and WASP-43\,b experience the strongest mass loss rates.
Our improved \lya{} flux estimates reveal that in seven of the nine systems expanded atmospheres could be detectable through \lya{} transit spectroscopy by stacking a small number of transit observations.
While WASP-80\,b, WASP-77\,b, and WASP-43\,b are good targets because of their strong predicted mass loss rates, WASP-8\,b is an interesting case because its higher mean density could locate the planet close to the transition from a strong photoevaporative wind to a stable atmosphere.
We find only weak indications for an angular momentum transfer from the orbiting hot Jupiters to the host stars in the two binary systems with expected strong tidal interactions.

We conclude that currently
the systems WASP-80, WASP-43, and WASP-77 represent the most promising candidates to search for absorption signals of the expanded atmospheres of hot Jupiters.

\begin{acknowledgements}
We (MS) acknowledge support through Verbundforschung (50OR 1105) and
the German National Science Foundation (DFG) within the Research Training College 1351.
PCS acknowledges support by the
DLR under 50 OR 1307 and by an ESA research fellowship.
This publication makes use of data products from the Two Micron All Sky Survey, which is a joint project of the University of Massachusetts and the Infrared Processing and Analysis Center/California Institute of Technology, funded by the National Aeronautics and Space Administration and the National Science Foundation.\\
This research has made use of the SIMBAD database,
operated at CDS, Strasbourg, France.This research has made use of the Exoplanet Orbit Database and the Exoplanet Data Explorer at exoplanets.org.
This paper makes use of data from the first public release of the WASP data (Butters et al. 2010) as provided by the WASP consortium and services at the NASA Exoplanet Archive, which is operated by the California Institute of Technology, under contract with the National Aeronautics and Space Administration under the Exoplanet Exploration Program.
\end{acknowledgements}


\bibliographystyle{aa}
\setlength{\bibsep}{0.0pt}
\bibliography{salz_hJ_hosts}

\end{document}